\documentclass[aps,floatfix,eqsecnum,showpacs,preprint,tightenlines,nofootinbib]{revtex4-1}
\usepackage{epsfig}

\usepackage{amsmath}
\usepackage{amssymb}
\usepackage{color}

\begin{document}



\title{Uncertainty of three-nucleon continuum observables arising from uncertainties of two-nucleon potential parameters}
\author{Yu. Volkotrub}
\affiliation{M. Smoluchowski Institute of Physics, Jagiellonian University, PL-30348 Krak\'ow, Poland}
\author{J.~Golak}
\affiliation{M. Smoluchowski Institute of Physics, Jagiellonian University, PL-30348 Krak\'ow, Poland}
\author{R.~Skibi{\'n}ski}
\affiliation{M. Smoluchowski Institute of Physics, Jagiellonian University, PL-30348 Krak\'ow, Poland}
\author{K. Topolnicki}
\affiliation{M. Smoluchowski Institute of Physics, Jagiellonian University, PL-30348 Krak\'ow, Poland}
\author{H.~Wita{\l}a}
\affiliation{M. Smoluchowski Institute of Physics, Jagiellonian University, PL-30348 Krak\'ow, Poland}
\author{E. Epelbaum}
\affiliation{Ruhr-Universit\"at Bochum, Fakult\"at f\"ur Physik und Astronomie, Institut f\"ur Theoretische Physik II, D-44780 Bochum, Germany}
\author{H. Krebs}
\affiliation{Ruhr-Universit\"at Bochum, Fakult\"at f\"ur Physik und Astronomie, Institut f\"ur Theoretische Physik II, D-44780 Bochum, Germany}
\author{P. Reinert}
\affiliation{Ruhr-Universit\"at Bochum, Fakult\"at f\"ur Physik und Astronomie, Institut f\"ur Theoretische Physik II, D-44780 Bochum, Germany}

\date{\today}

\begin{abstract}
Propagation of uncertainties from two-nucleon potential parameters to three-nucleon observables, that is statistical errors
for the neutron-deuteron elastic scattering and the deuteron breakup reaction at neutron laboratory energies up to 200 MeV is investigated. 
{To that} end we use the chiral nucleon-nucleon interaction with the semi-local momentum-space 
regularization at various orders of the chiral expansion, exploiting knowledge of the covariance matrix 
of its parameters. 
For both reactions we compare statistical uncertainties for chiral predictions with the uncertainties obtained in the same way
but for the semi-phenomenological One-Pion-Exchange two-nucleon force.
In addition for the elastic scattering we show also the truncation errors 
arising from restriction to a given order of chiral predictions, estimated among others within the Bayesian method,
and the cutoff dependence of chiral predictions.
We find that the resulting statistical uncertainty is smaller than the truncation 
errors for the chiral force at lower orders of the chiral expansion. 
At the higher orders of the chiral expansion and at low energies the statistical errors exceed the truncation ones
but at intermediate and higher energies truncation errors are more important. 
Overall, magnitudes of the theoretical uncertainties are small and amount up to 0.5\%-4\%, depending on 
the observable and energy.
We also find that the magnitudes of statistical uncertainties for the chiral and semi-phenomenological potentials are similar
and that the dependence of predictions on the regularization parameter values is important at all investigated energies. 
\end{abstract}


\maketitle


\section{Introduction}
The interactions between nucleons originate from interactions between quarks and gluons in the nonperturbative 
regime of Quantum Chromodynamics (QCD).  Since currently 
the nuclear forces derived directly from QCD are not available,
various effective models of nuclear interactions are used. In recent decades there has been 
increased interest in potentials based on the Chiral Effective Field Theory ($\chi$EFT), 
linked to QCD and its symmetries. Within this approach it is possible to construct a consistent effective Hamiltonian 
with two- and many-body nuclear forces which incorporate all possible contributions up to 
a given order of the chiral expansion. Commonly nucleons and pions are chosen as relevant degrees 
of freedom, see Refs.~\cite{Epelbaum_review, Epelbaum2012, Machleidt_review, Epelbaum2-arxiv, Piarulli2020} for 
more information on various chiral interactions.

Parallel to the development of the nuclear force models, the question how to estimate uncertainties 
of the calculated nuclear observables within a given model has arisen
and various ideas for estimating theoretical uncertainties within the $\chi$EFT framework, 
see Refs.~\cite{Epelbaum_review, imp1, imp2, lenpic3, lenpic4, Reinert, Epelbaum2019, Epelbaum1-arxiv, 
Epelbaum2-arxiv, Furnstahl, Melendez, Ekstrom, special_issue} have been proposed and discussed. 
In the past, the uncertainty quantification of theoretical predictions in nuclear physics was treated with 
less care compared to the error analysis of experimental results. The estimation of uncertainties of 
theoretical predictions in few-nucleon systems was based mainly on comparison of predictions obtained using 
various models of the nuclear interaction. Such models describe, in \textit{ab initio} calculations, low-energy observables with relatively high precision~\cite{Kalantar_Nogga}. The CD-Bonn~\cite{Mac01}, the AV18~\cite{AV18} or the chiral interactions derived by the Bochum-Bonn~\cite{imp1, imp2, Reinert}, the Moscow (Idaho)-Salamanca~\cite{Entem_2017} or the 
Livermore~\cite{Piarulli1, Piarulli2} groups, are good examples of such forces. Each of these models has some number of free parameters, 
whose values are fixed by the data. In the past, the authors of interaction models usually restricted themselves to determination of the values of the parameters 
but skipped their errors analysis, see for example Ref.~\cite{AV18}. The situation has changed since
the Granada group revised the existing database for the nucleon-nucleon (NN) scattering and derived based on this data set 
the One-Pion-Exchange (OPE) Gaussian NN force model, Refs.~\cite{Granada, Navarro2014} and other potentials~\cite{Navarro2015}. 
The careful statistical treatment applied during the fitting procedure allowed authors of Refs.~\cite{Navarro2014, Navarro2015} to obtain the covariance matrix of the potential parameters. Using it, we studied the propagation of the uncertainties of the OPE-Gaussian potential parameters from the two-nucleon (2N) system to the 
elastic neutron-deuteron (nd) scattering observables in Ref.~\cite{Skibinski_2018}, determining for the first time in a quantitative way the corresponding theoretical uncertainties (called statistical uncertainties in the following). We refer the reader to Ref.~\cite{Skibinski_2018} for a more 
general discussion on various types of theoretical uncertainties for the elastic nd scattering observables 
and to a special issue of \textit{Journal of Physics G: Nuclear and Particle Physics}~\cite{special_issue} for other observables and processes. 

In this paper we show results for the elastic nd scattering and the neutron-induced deuteron 
breakup process obtained with the newest $\chi$EFT family of potentials from the Bochum group~\cite{Reinert}. 
For this interaction, derived completely up to the fifth order of the perturbative 
expansion (N$^{4}$LO), the semi-local regularization in momentum space (SMS) has been applied. 
Further, for this potential the covariance matrix of its free parameters (obtained with the 
Granada database~\cite{Navarro2014}) is known, which allows us to study, for the first time for a chiral force, the propagation 
of uncertainties of NN interaction parameters to three-nucleon (3N) continuum observables.  
Also the dependence of the uncertainty pattern on the order of the chiral expansion and on the regulator value 
is additionally studied.

On top of the statistical uncertainties, also the so-called truncation errors, which are 
uncertainties arising from restriction 
to a given order of the chiral expansion can be evaluated. This was done for the first time in Ref.~\cite{imp1},
where a simple prescription to estimate the truncation errors for the NN system was proposed. 
This prescription has been extended to many-nucleon systems in Ref.~\cite{lenpic3}. 
Though simple the algorithm of~\cite{imp1} does not give a statistical interpretation of truncation errors. 
Those can be estimated within Bayesian methods, see for example Refs.~\cite{Furnstahl, Melendez} focused 
on NN observables. Recent papers~\cite{Epelbaum1-arxiv, Epelbaum2-arxiv} have presented Bayesian results 
for truncation errors for observables in neutron-deuteron scattering below the pion production threshold. 
In the present paper we employ the Bayesian approach of Ref.~\cite{Epelbaum1-arxiv} and compare resulting 
truncation errors to those obtained within the method from Ref.~\cite{lenpic3} as well as to the uncertainty due to the regulator 
dependence and the statistical errors.

This paper is organized as follows. In Sec.~\ref{sec:FaddeevFormalism} we outline the Faddeev formalism 
for 3N calculations. In Sec.~\ref{sec:Determination} we briefly describe our method used to estimate 
the propagation of the uncertainties of the potential parameters from the 2N system to the elastic 3N scattering observables. 
Sections~\ref{sec:ResultElastic} and~\ref{sec:ResultBreakup} describe results for elastic scattering 
and breakup reactions, respectively. Specifically, we discuss the theoretical statistical uncertainties and compare them with the truncation errors for a few chosen observables. We summarize in Sec.~\ref{sec:Summary}.

\section{\label{sec:FaddeevFormalism}Formalism for 3N scattering}

The nucleon-deuteron scattering observables can be obtained using the formalism of the 3N Faddeev equation. 
This is one of the standard techniques to investigate 3N reactions and has been described in detail many 
times, see for example Refs.~\cite{Glockle_book, Glockle_raport, Witala}. Thus we only briefly describe the key 
steps of this approach. The starting point for 3N calculations is solving the Lippmann-Schwinger equation 
with a given NN interaction $V$ to get the NN $t$ operator:
\begin{equation}
t=V + V \tilde{G_0} t\;,
\label{eq_LS}
\end{equation}
where $\tilde{G_0}$ is the free propagator of two nucleons. The  $t$ operator enters the 3N Faddeev scattering 
equation which, neglecting the 3N force, is written as:
\begin{equation}
T \vert \phi \rangle = tP \vert \phi \rangle + tPG_0 T \vert \phi \rangle\; .
\label{eq_Fadd_NN_3N}
\end{equation}
Here the initial state $\vert \phi \rangle$ is composed of the deuteron wave function and the momentum eigenstate 
of the projectile nucleon, $G_{0}$ is the free 3N propagator and $P$ is the permutation operator 
$P \equiv P_{12}P_{23} + P_{13}P_{23}$ built from transpositions $P_{ij}$, which interchange 
particles $i$ and $j$. Next the transition amplitudes, $U$ for elastic Nd scattering 
and $U_{0}$ for the deuteron breakup process, are calculated via 
\begin{equation}
\begin{split}
\langle \phi^{'}\vert U \vert \phi \rangle &= \langle \phi^{'}\vert PG^{-1}_{0} \vert \phi \rangle + \langle \phi^{'}\vert PT \vert \phi \rangle\;, \\
\langle \phi^{'}\vert U_{0} \vert \phi \rangle&= \langle \phi^{'}\vert (1 + P)T \vert \phi \rangle
\end{split}
\label{eq_transition}
\end{equation}
and used to compute 3N scattering observables in the standard way~\cite{Glockle_raport}. 
$\vert \phi$'$ \rangle$ in~Eq.~(\ref{eq_transition}) denotes the suitable final two-body (nd)
or three-body breakup state. In the latter case $\vert \phi$'$ \rangle$ is a product 
of two relative-momentum eigenstates describing free motion of three outgoing nucleons. 

In practice we work in the momentum-space 
partial wave basis $\vert p, q, \alpha\rangle$, where $p \equiv \vert \vec{p}\vert$ and $q \equiv \vert \vec{q}\vert$ are the magnitudes of the Jacobi momenta $\vec{p}$ and $\vec{q}$; $\alpha$ represents a set of discrete quantum numbers for the 3N system in the $jI$-coupling, and is defined as $\alpha \equiv \vert (ls)j; (\lambda\frac{1}{2})I;(jI)JM_{J}; (t\frac{1}{2})TM_{T}\rangle$. Here $l, s, j$ and $t$ are the orbital angular momentum, total
spin, total angular momentum, and total isospin of the $(2-3)$
subsystem. Further, $\lambda$ is the orbital angular momentum of nucleon $1$, which together with its spin $\frac{1}{2}$, couples to the total angular 
momentum $I$ of nucleon 1. The angular momenta $j$ and $I$ couple to the total angular momentum of the 3N system $J$, and $M_{J}$ denotes its projection 
on the quantization axis. The quantum numbers $T$ and $M_{T}$ describe the total isospin of the 3N system and its third component, respectively.
Equation~(\ref{eq_Fadd_NN_3N}) is solved numerically by generating its Neumann series, which is subsequently summed up using the Pad\`e method. 
For the investigations presented here we use all partial waves with $j \leq 5$ and $J \leq \frac{25}{2}$, which is sufficient 
to guarantee convergence of our predictions at the considered energies~\cite{Glockle_raport}.

\section{\label{sec:Determination}Determination of the statistical uncertainties in the 3N system}
\label{Sec_Determination}

Computation of the above-defined statistical uncertainties for a specific observable requires a big sample of predictions 
obtained with different sets of parameters within a given model of the NN interaction. Prerequisite is the knowledge 
of the covariance matrix (or equivalently the correlation matrix) of the NN potential parameters, 
as is the case for the semilocally regularized in momentum space (SMS) chiral potential of Ref.~\cite{Reinert}. 
We apply here the same method as was 
used previously in Ref.~\cite{Skibinski_2018} to study the propagation of the uncertainties 
of the OPE-Gaussian potential parameters from the 2N system to 3N observables in elastic 
neutron-deuteron scattering. Therefore, we only briefly describe our algorithm to determine 
the statistical uncertainty and use it in the following for the chiral SMS force. 
Namely, given the expectation values (this set of potential parameters we call $S_0$ in the following), 
and correlation coefficients for the potential parameters, 
we sample, from the multivariate normal distribution, 50 sets of the potential parameters. 
For each set, we solve Eqs.~(\ref{eq_Fadd_NN_3N})--(\ref{eq_transition}) and compute 3N observables.
Various possible estimators of the statistical uncertainties have been described in Ref.~\cite{Skibinski_2018} 
and compared with each other. We use $\frac{1}{2}\Delta_{68\%}$\footnote{$\Delta_{68\%}$ is the spread of 
results in the set of 34 (68\% of 50) predictions based on different sets of the NN potential parameters. 
The set of 34 observables is constructed by discarding the 8 lowest and the 8 highest predictions 
for a given observable and at specific scattering angle and energy.} as an optimal measure 
for dispersion of predictions and consequently as an estimator of the statistical uncertainty at a
given energy and a scattering angle. The same method was used to quantify 
the statistical error of the $^{3}$H binding energy in Ref.~\cite{Navarro_H} and to estimate the uncertainties 
of the $^{4}$He bound states in Ref.~\cite{Navarro_He}. 
Note that in the case of the SMS potential 
the regulator dependence and the availability of predictions at different orders of the chiral expansion
increase the required number of computations substantially.

\section{\label{sec:ResultElastic}Results for the elastic \boldmath$\rm Nd$ scattering}
\label{Sec_Ndscattering}
\begin{figure}[ht]
\includegraphics[width=1\textwidth,clip=true]{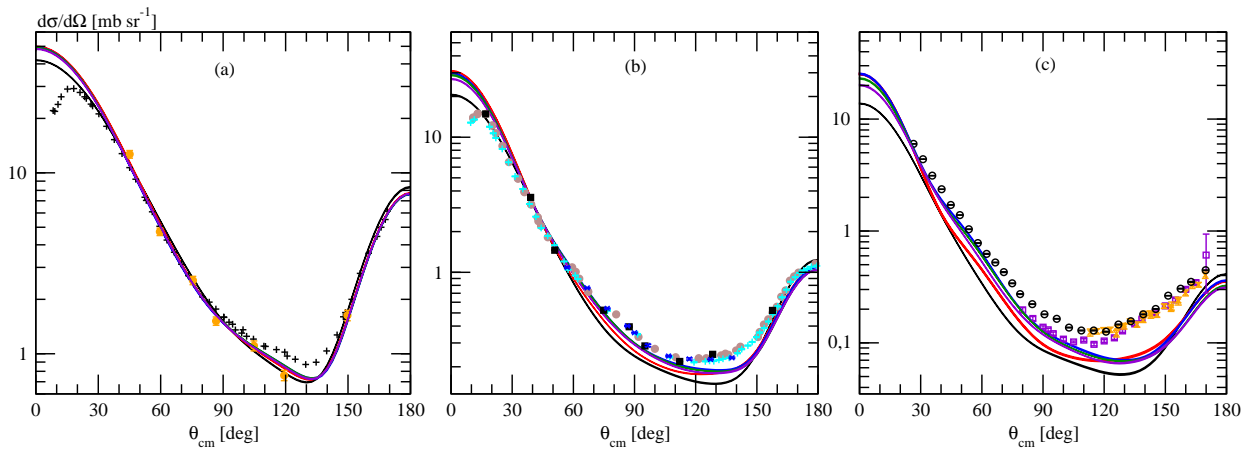}
\caption{(color online) The differential cross section d$\sigma$/d$\Omega$ for the elastic nd scattering process at the incoming neutron laboratory  energy (a) $E = 65$ MeV, (b) $E = 135$ MeV and (c) $E = 200$ MeV as a function of the center-of-mass scattering angle $\theta_{c.m.}$. The black, red, blue, green and violet bands represent statistical uncertainties based on the chiral NLO, N$^{2}$LO, N$^{3}$LO, N$^{4}$LO and N$^{4}$LO$^+$ ($\Lambda$~=~450~MeV) SMS potentials, respectively.
The experimental data are in: (a) from Ref.~\cite{Shimizu} ($pd$ pluses) and~\cite{Ruhl} (nd orange circles), (b) from Ref.~\cite{dcs_dp135_kimiko} ($dp$ brown circles),~Ref.~\cite{dcs_riken2} ($dp$ cyan pluses), Ref.~\cite{dcs_kimiko} ($pd$, $E = 135$ MeV, blue $\times$'s) and~Ref.~\cite{dcs_kimiko} ($pd$ black squares), and in (c) from Ref.~\cite{Adelberger} ($pd$ violet squares, $E$~=~198~MeV), Ref.~\cite{Igo} ($pd$ orange $\times$'s, $E$~=~180~MeV), and Ref.~\cite{Ermisch} ($pd$ black circles, $E$~=~198~MeV).}
\label{fig1} 
\end{figure}

We start presenting our results from discussing the
dependence of statistical uncertainties, obtained with the chiral 
SMS NN interaction~\cite{Reinert}, on the order of the chiral expansion. This is done 
for selected observables in elastic neutron-deuteron (nd) scattering at 
three laboratory energies of the incoming neutron: $E =$ 65, 135 and 200~MeV. 
We employ the regularization parameter $\Lambda$~=~450~MeV. In Fig.~\ref{fig1} we show the quality   
of the elastic scattering cross section data description obtained with the SMS chiral force. Our predictions are represented by 
bands which for each order of the chiral expansion cover a $\Delta_{68\%}$ estimator 
of the statistical uncertainty range.
At the lowest energy, $E=$65~MeV, predictions are very close to one another except for the 
NLO, which separates clearly at forward and backward scattering angles. The narrowness of bands 
clearly shows that at this energy the uncertainty of the nd elastic cross section arising 
from the uncertainty of the NN potential parameters is very small for all scattering angles. 
At two higher energies spreads of the different order of chiral expansion results become larger, 
however, the values of statistical uncertainties remain small. This is similar to the results for the OPE-Gaussian 
force~\cite{Skibinski_2018}, where small values of statistical uncertainties have been found for elastic nd scattering observables.
The observed discrepancy with the proton-deuteron cross section data at small scattering angles is well understood 
as a result of neglecting the Coulomb force in our nd calculations~\cite{Deltuva}. 
The discrepancy around the minimum of the cross section is due to omitting 3N force contributions.

\begin{figure}[ht]
\includegraphics[width=1\textwidth,clip=true]{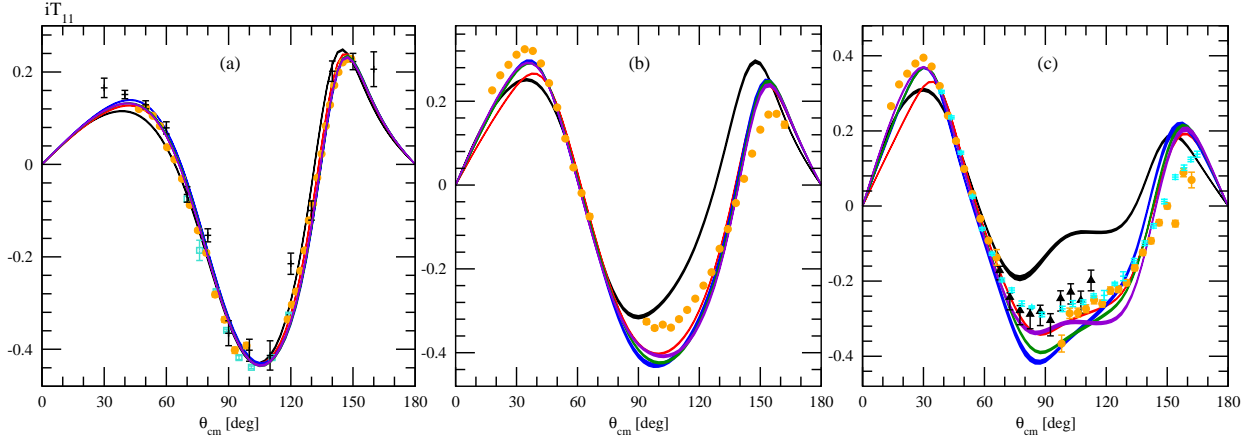}
\caption{(color online) The deuteron vector analyzing power iT$_{11}$ for elastic nd scattering at the same energies as in Fig.~\ref{fig1} shown as a function of the scattering angle $\theta_{c.m.}$. For description of bands see Fig.~\ref{fig1}.
The data are in: (a) from Ref.~\cite{Witala_it11} ($pd$ pluses), Ref.~\cite{Stephan} ($pd$ orange circles), and Ref.~\cite{Mardanpour} ($pd$ cyan squares), (b) from Ref.~\cite{Przewoski} ($pd$ orange circles), and (c) from Ref.~\cite{Przewoski} ($pd$ orange circles), Ref.~\cite{Cadman} ($pd$ black up-triangles, $E$~=~197~MeV), Ref.~\cite{Sekiguchi} ($pd$ cyan pluses, $E$~=~186.6~MeV).}
\label{fig2} 
\end{figure}

\begin{figure}[h]
\includegraphics[width=1\textwidth,clip=true]{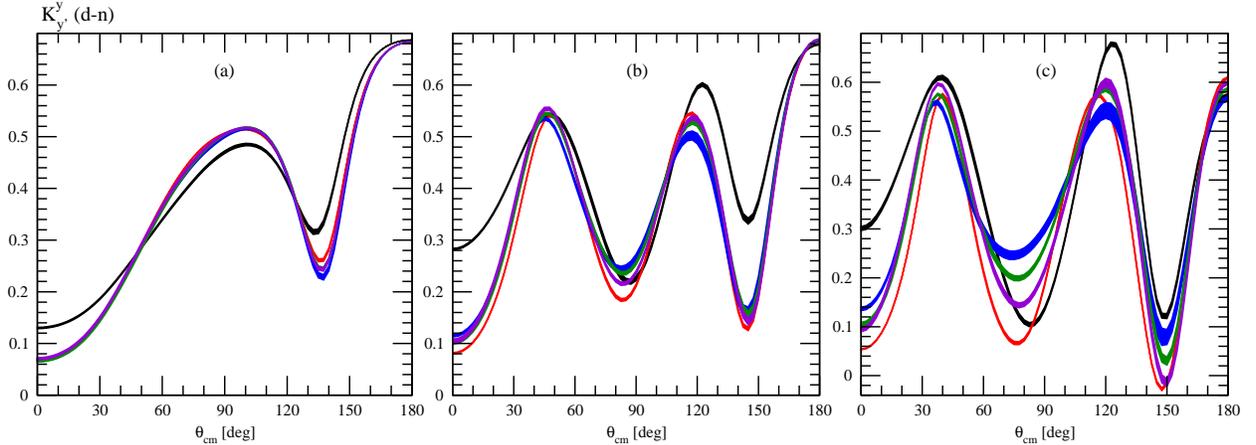}
\caption{(color online) The deuteron to neutron spin-transfer coefficient $K^{y}_{y'}$(d-n) at the same energies as in Fig.~\ref{fig1} shown as a function of the scattering angle $\theta_{c.m.}$. For description of bands see Fig.~\ref{fig1}.}
\label{fig3} 
\end{figure}

\begin{figure}[ht]
\includegraphics[width=1\textwidth,clip=true]{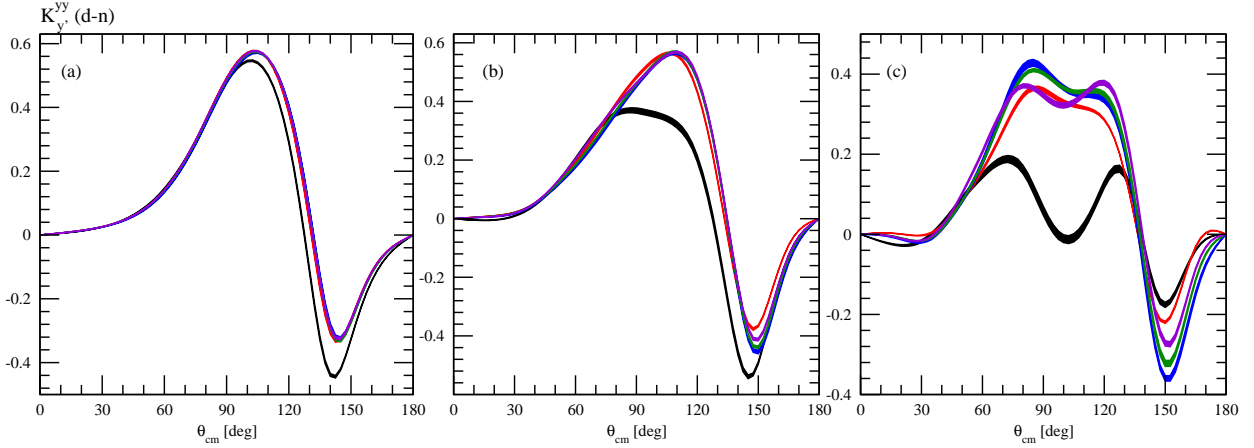}
\caption{(color online) The deuteron to neutron spin-transfer coefficient $K^{yy}_{y'}$(d-n) at the same energies as in Fig.~\ref{fig1} as a function of the scattering angle $\theta_{c.m.}$. Bands are as in Fig.~\ref{fig1}.}
\label{fig4} 
\end{figure}

In Fig.~\ref{fig2} we show the deuteron vector analyzing power iT$_{11}$. In this case, the chiral SMS interaction at the NLO order of chiral 
expansion fails to describe data at both higher energies. The statistical uncertainties remain small 
for all energies and orders and are negligible compared to differences between different order predictions at 200~MeV.
In Fig.~\ref{fig3} we show the deuteron to neutron spin-transfer coefficient $K^{y}_{y'}$(d-n), for which the differences between 
predictions at various orders of the chiral expansion are especially big at 200~MeV. 
The statistical uncertainty obtained 
with the chiral N$^{2}$LO force is relatively small at all energies and slightly grows with the increasing energy. 
For example, the difference between the two predictions obtained with the chiral N$^{2}$LO and N$^{4}$LO SMS potentials amounts $\approx 60$\% at the minimum for 200 MeV at $\theta_{c.m.} = 147.5^{\circ}$, while the N$^{2}$LO (N$^{4}$LO) statistical uncertainties reach 0.27\% (0.88\%). In the case of the deuteron to neutron spin-transfer coefficient $K^{yy}_{y'}$(d-n) shown in Fig.~\ref{fig4} we do not see such large statistical errors as in the case of $K^{y}_{y'}$(d-n), but still their magnitude changes with the energy. Actually, we observe the following behavior: for the chiral SMS N$^{2}$LO interaction the statistical uncertainty increases at $E =$ 135 MeV compared to the $E =$ 65 MeV, but at $E =$ 200 MeV the statistical uncertainty decreases in the range of $\theta_{c.m.} \in  [72.5^{\circ}, 150^{\circ}]$ compared to the $E =$ 135 MeV case. For the chiral SMS N$^{4}$LO interaction the statistical uncertainty decreases at $\theta_{c.m.} \in [40^{\circ}, 62.5^{\circ}] \cup [92.5^{\circ}, 112.5^{\circ}]\cup [145^{\circ}, 180^{\circ}]$ at $E =$ 135 MeV compared to the energy 65 MeV and its magnitude further 
decreases at $\theta_{c.m.} \in [0^{\circ}, 47.5^{\circ}] \cup [92.5^{\circ}, 130^{\circ}]$  at $E =$ 200 MeV compared to results at $E =$~135~MeV. 
To quantify this behaviour we give example at $\theta_{c.m.} = 145^{\circ}$ where the $\frac12 \Delta_{68\%}$ reaches 0.22\%(0.24\%), 0.30\%(0.43\%), and
0.27\%(0.74\%) of $K^{yy}_{y'}$(d-n) N$^2$LO(N$^4$LO) predictions at $E$=65 MeV, 135 MeV, and 200 MeV, respectively.

It is interesting to compare magnitudes of the statistical errors with other kinds of theoretical uncertainties. 
Here, we would like to focus on the truncation errors present intrinsically in the chiral approach. 
Using the method from Ref.~\cite{lenpic3} we calculate the truncation error of a given 3N observable and compare its size with 
the statistical uncertainties already obtained.
Namely, any 3N scattering observable $X$ at a fixed cutoff value can be expanded
up to the i-th order of the chiral expansion 
($i = 0, 2, 3, \ldots$) in the form
\begin{align}
X = X^{(0)} + \Delta X^{(2)} + \Delta X^{(3)} + \ldots + \Delta X^{(i)}
\label{eq_truncation0}
\end{align}
Then the truncation error $\delta(X)^{(i)}$ of an observable $X$ at the $i$-th order of the chiral expansion with $i = 0, 2, 3, \ldots$, is~\cite{lenpic3}
\begin{align}
\begin{split}
\delta(X)^{(0)} &\geq max (Q^{2}\vert X^{0} \vert, \vert X^{(i \geq 0)} - X^{(j \geq 0)} \vert ), \\
\delta(X)^{(2)} & \geq max(Q^{3}\vert X^{0} \vert, Q \vert \Delta X^{(2)}, \vert X^{(i \geq 2)} - X^{(j \geq 2)} \vert), \\
\delta(X)^{(i)} &\geq max(Q^{i+1}\vert X^{0} \vert, Q^{i-1} \vert \Delta X^{(2)}, Q^{i-2} \vert \Delta X^{(3)} \vert), i \geq 3
\end{split}
\label{eq_truncation}
\end{align}
where $X^{(i)}$ denotes a prediction for the observable $X$ at $i$-th order, $\Delta X^{(2)} \equiv X^{(2)} - X^{(0)}$ 
and $\Delta X^{(i)} \equiv X^{(i)} - X^{(i-1)}$ for $i \geq 3$. Further additional conditions 
$\delta(X)^{(2)} \geq Q\delta(X)^{(0)}$ and $\delta(X)^{(i)} \geq Q\delta(X)^{(i-1)}$ 
for $i \geq 3$ are imposed on the truncation errors. 
Such estimation of truncation errors accounts for the fact that the 3N force is 
neglected in the current investigation. 

\begin{figure}[ht]
\includegraphics[width=1\textwidth,clip=true]{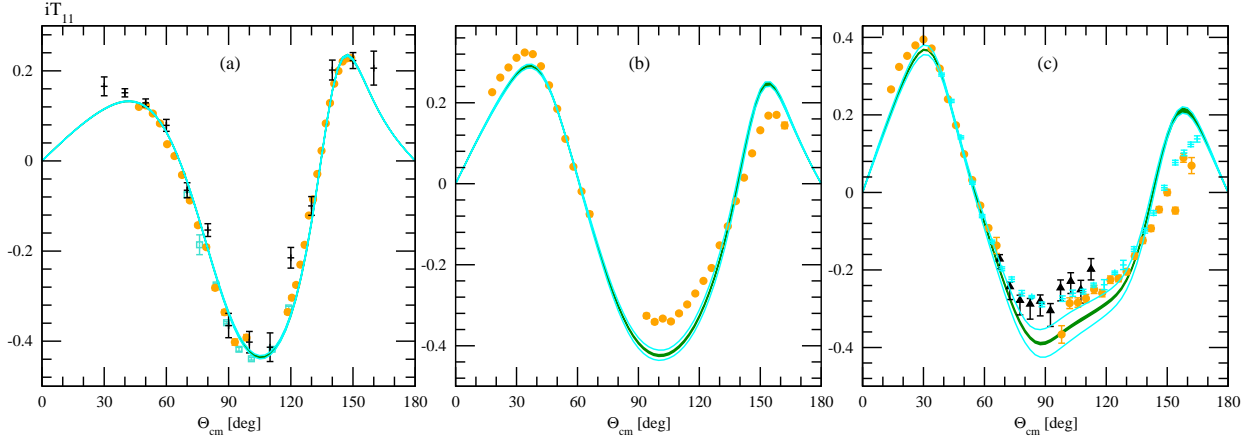}
\caption{(color online) The deuteron vector analyzing power i$T_{11}$ for elastic nd scattering at 
the same energies as in Fig.~\ref{fig1} as a function of the scattering angle $\theta_{c.m.}$. 
The green solid band represents the statistical uncertainties based on the chiral N$^{4}$LO ($\Lambda$~=~450~MeV) potential 
and the cyan lines represent the borders of the band for the truncation error for the same potential estimated using Eq.(\ref{eq_truncation}). The experimental data are the same as in Fig~\ref{fig2}.}
\label{fig5} 
\end{figure}

\begin{figure}[ht]
\includegraphics[width=1\textwidth,clip=true]{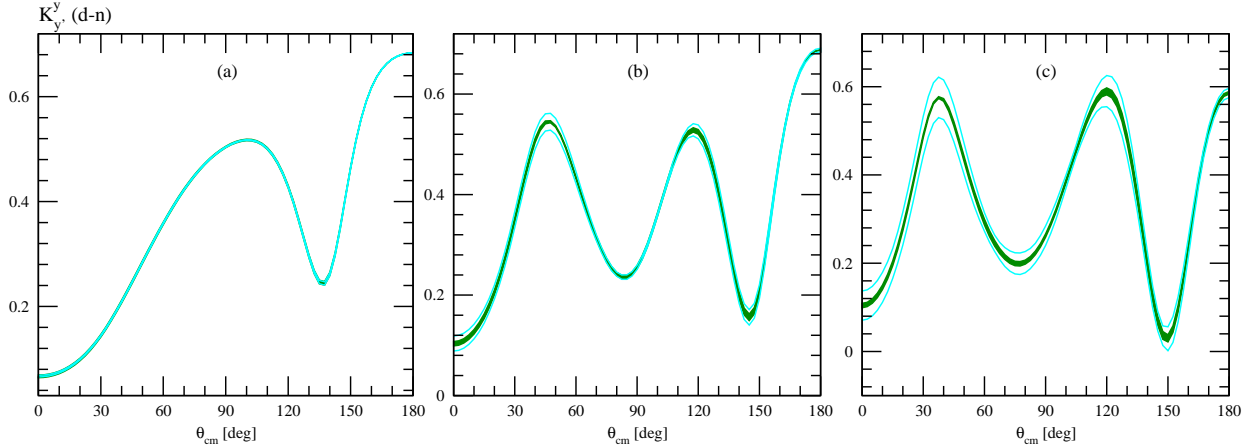}
\caption{(color online) 
The deuteron to neutron spin-transfer coefficient $K^{y}_{y'}$(d-n) for elastic nd scattering at the same energies as used in Fig.~\ref{fig1} as a function of the scattering angle $\theta_{c.m.}$. The band and lines are as in Fig.~\ref{fig5}.}
\label{fig6} 
\end{figure}

In Figs.~\ref{fig5} and~\ref{fig6} we show a comparison of the statistical and truncation errors for 
the deuteron vector analyzing power iT$_{11}$ and the deuteron to neutron spin-transfer coefficient $K^{y}_{y'}$(d-n). The N$^{4}$LO SMS interaction with $\Lambda =$~450 MeV is used and the same energies are taken as in Fig.~\ref{fig1}. For the sake of clarity, for the truncation errors we show only, with the blue curves, borders of the corresponding band. 

\begin{table}
\begin{tabular}{|l|l|r|c|c|c|c|}
\hline
$E$ [MeV] & $\theta_{c.m.}$ [deg] & iT$_{11}$($S_{0}$)& $\vert$ iT$_{11}$($S_{0}$) - iT$^{min}_{11} \vert$ & $\vert$ iT$^{max}_{11}$ - iT$_{11}$($S_{0})\vert$& $\frac{1}{2}\Delta_{68\%}$ & $ \delta$(iT$_{11}$)$^{(5)}$ \\ \hline
65        & 30                    & 0.115234       &0.000644 &0.000493       & 0.000569       & 0.000425        \\ \hline
          & 75                    & -0.117815      &0.000820 &0.000577       & 0.000699       & 0.001685        \\ \hline
          & 120                   & -0.342291      &0.002194 &0.001620       & 0.001911       & 0.002680        \\ \hline
          & 165                   & -0.089323      &0.000486 &0.000313       & 0.000399       & 0.000599        \\ \hline
135       & 30                    & 0.270409       &0.001503 &0.001204       & 0.001354       & 0.004570         \\ \hline
          & 75                    & -0.233802      &0.001052 &0.001486       & 0.001269       & 0.009550         \\ \hline
          & 120                   & -0.326824      &0.003196 &0.001554       & 0.002375       & 0.015205        \\ \hline
          & 165                   & 0.155717       &0.001594 &0.001174       & 0.001385       & 0.003405        \\ \hline
200       & 30                    & 0.367730       &0.000706 &0.001307       & 0.001007       & 0.012490        \\ \hline
          & 75                    & -0.307313      &0.002202 &0.003750       & 0.002976       & 0.026585        \\ \hline
          & 120                   & -0.286319      &0.003595 &0.002643       & 0.003119       & 0.028870        \\ \hline
          & 165                   & 0.175372       &0.003484 &0.002954       & 0.003219       & 0.007540        \\ \hline
\end{tabular}
\caption{The deuteron analyzing power iT$_{11}$ at given incoming neutron energy $E$ 
and scattering angle $\theta_{c.m.}$, for the expectation values of the chiral SMS N$^{4}$LO potential parameters (denoted as set $S_{0}$), 
and its statistical $\frac12 \Delta_{68\%}$ as well as truncation $\delta^{(5)}$ errors. In addition, the borders of iT$_{11}$ 
for 34 sets (iT$^{min}_{11}$ and iT$^{max}_{11}$) are given. $\Delta_{68\%} \equiv iT^{max}_{11} - iT^{min}_{11}$.}
\label{tab1}
\end{table}

\begin{figure}
\includegraphics[width=5.3cm]{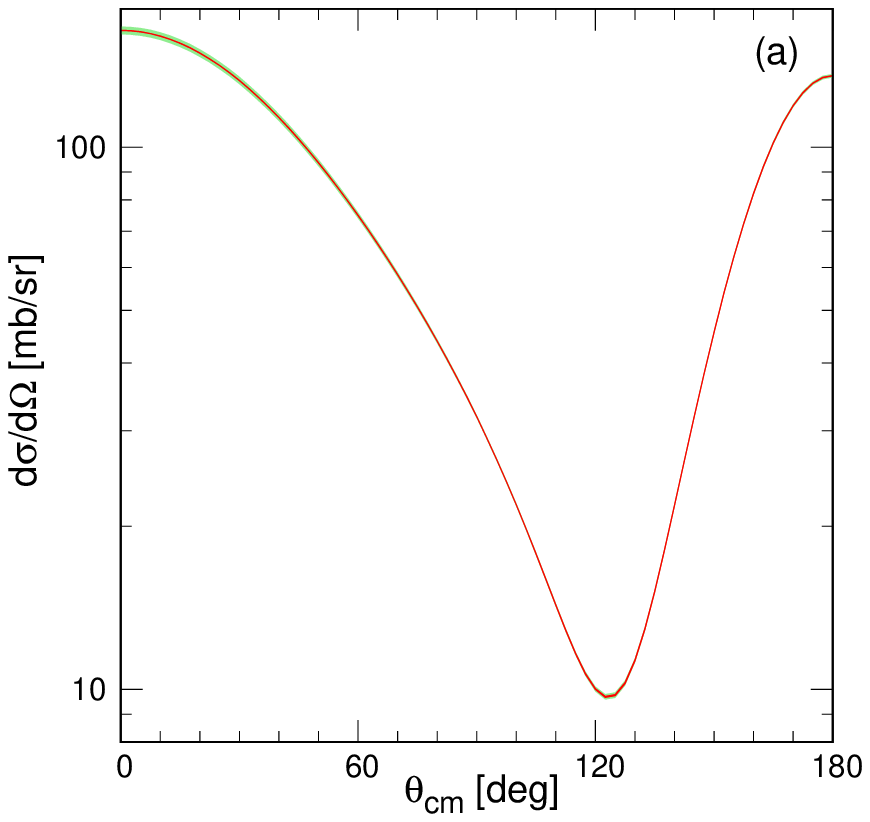}
\includegraphics[width=5.3cm]{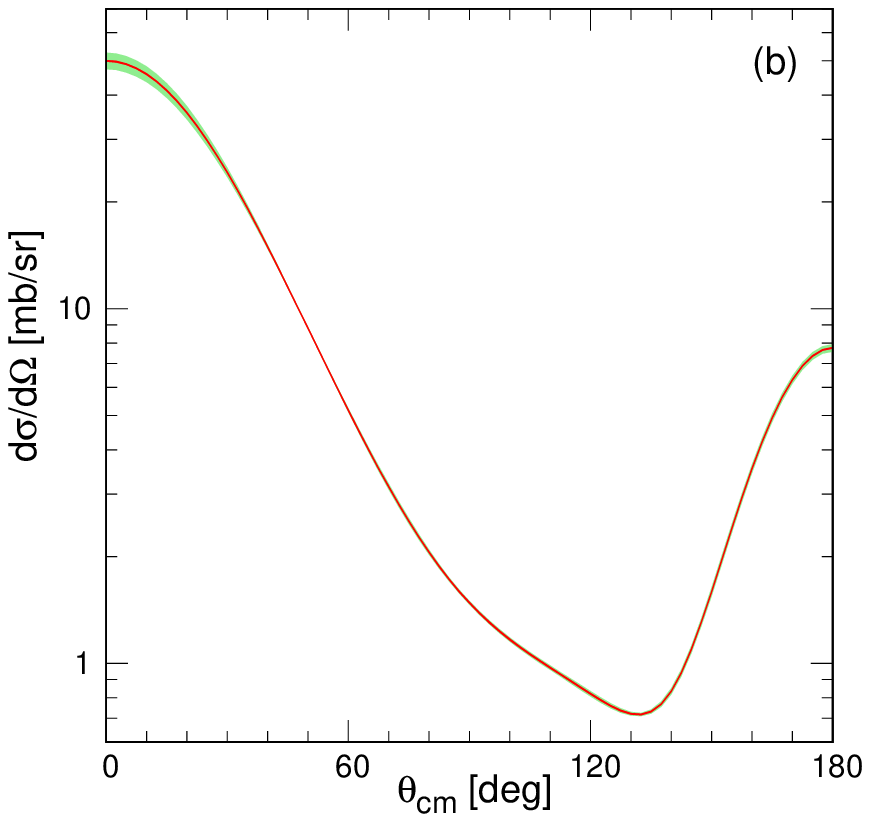}
\includegraphics[width=5.3cm]{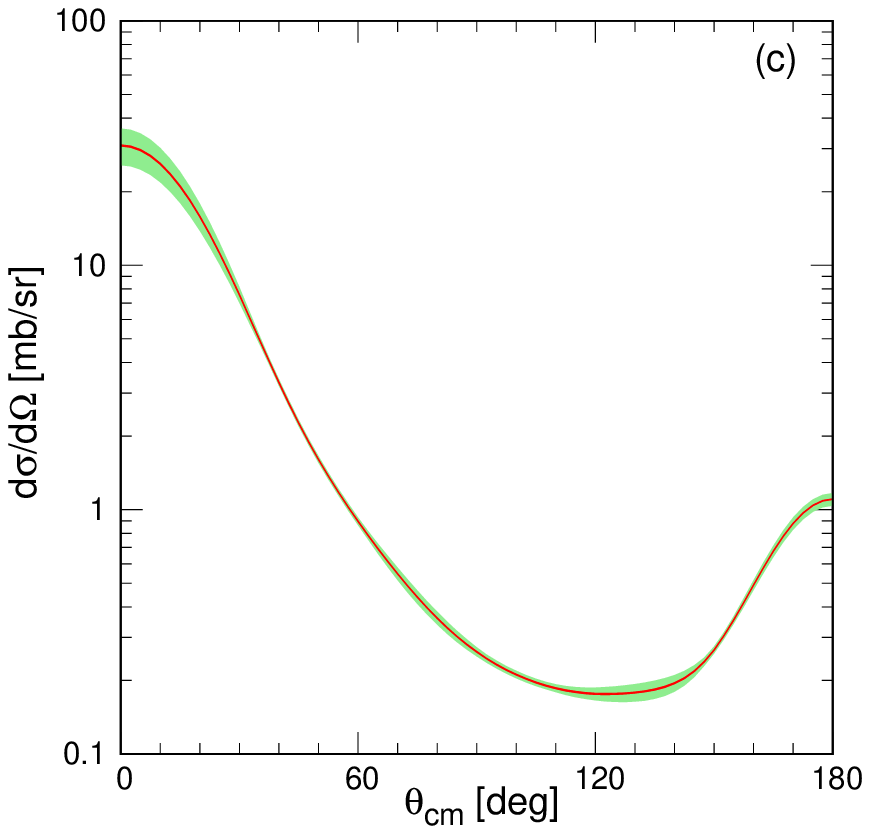}
\includegraphics[width=5.3cm]{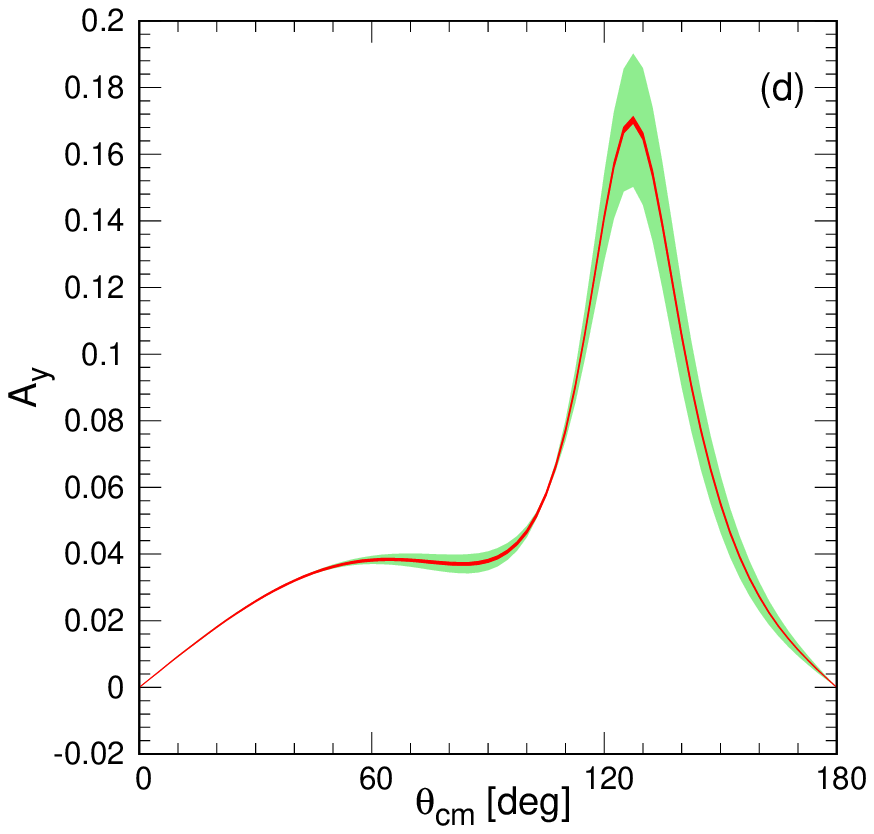}
\includegraphics[width=5.3cm]{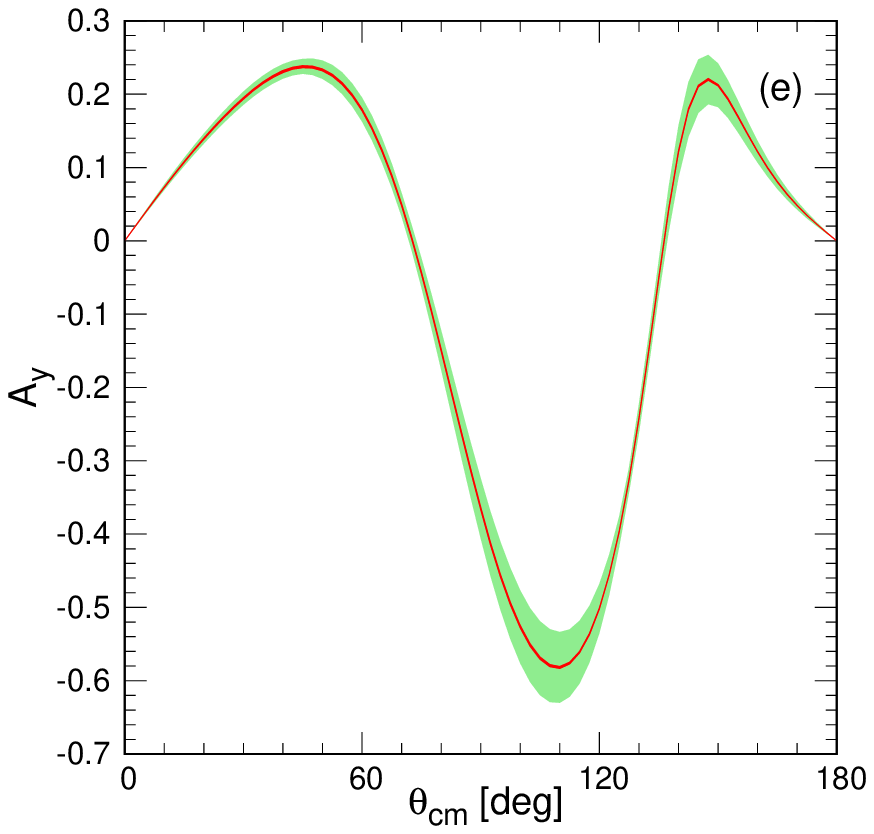}
\includegraphics[width=5.3cm]{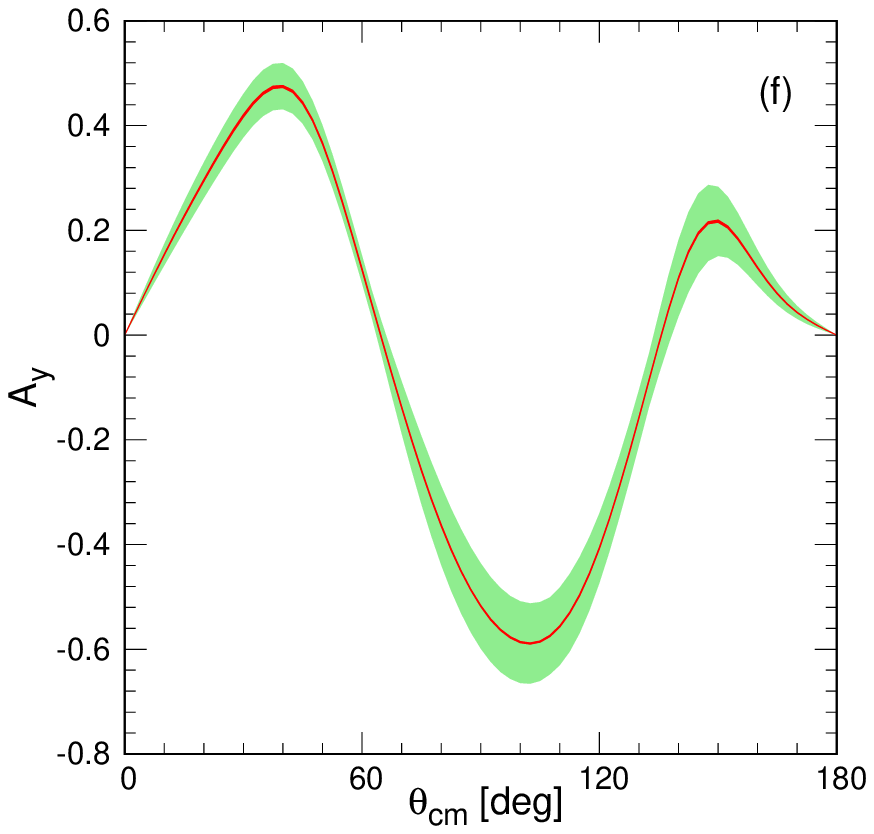}
\caption{(color online) 
Predictions for the differential cross section d$\sigma$/d$\Omega$ 
(top panels (a), (b) and (c)) and the neutron analyzing power $A_{y}$ 
(bottom panels (d), (e) and (f)) 
for elastic nd scattering at the incoming neutron laboratory energy
(a), (d) $E = 13$ MeV, (b), (e) $E = 65$ MeV and (c), (f) $E = 135$ MeV 
as a function of the center-of-mass scattering angle $\theta_{\rm{c.m.}}$. 
The red and light-green bands denote the statistical uncertainty and 68\% DoB interval using the Bayesian model $\bar{C}^{650}_{0.5-10}$ based on the chiral SMS N$^{2}$LO ($\Lambda =$~450~MeV) NN potential, respectively. }
\label{bayes_stat_N2LO}
\end{figure}

\begin{figure}
\includegraphics[width=5.3cm]{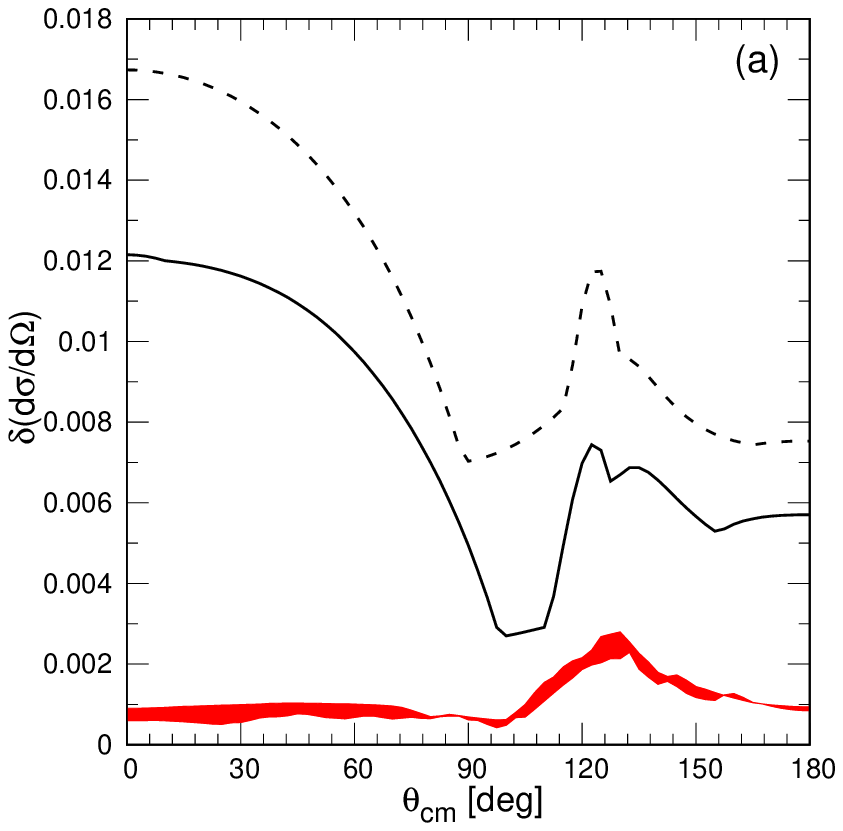}
\includegraphics[width=5.3cm]{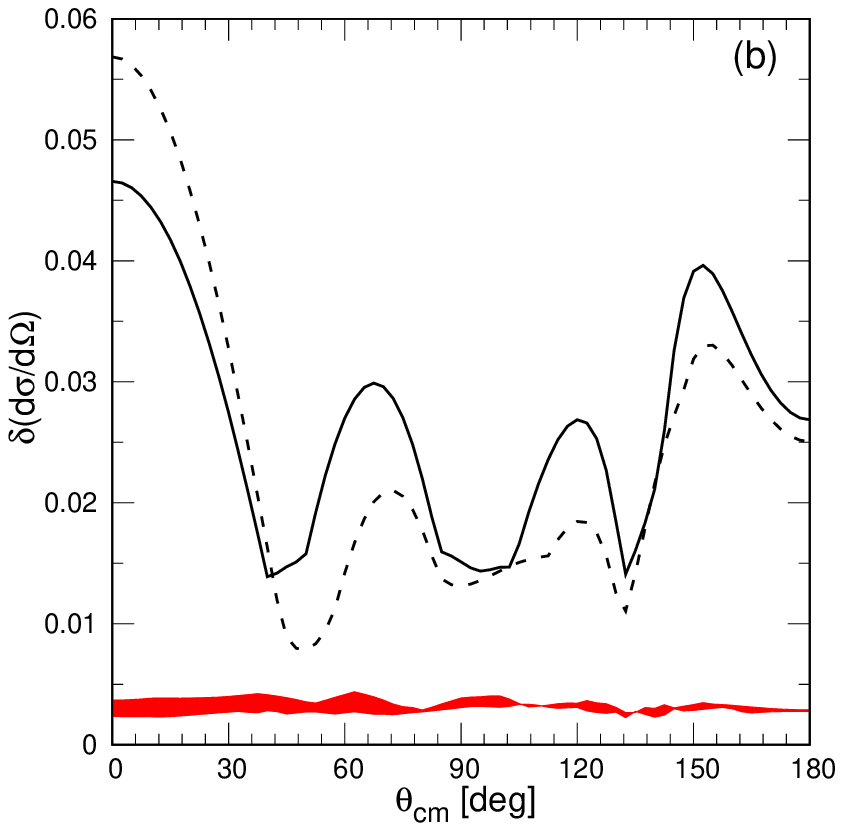}
\includegraphics[width=5.3cm]{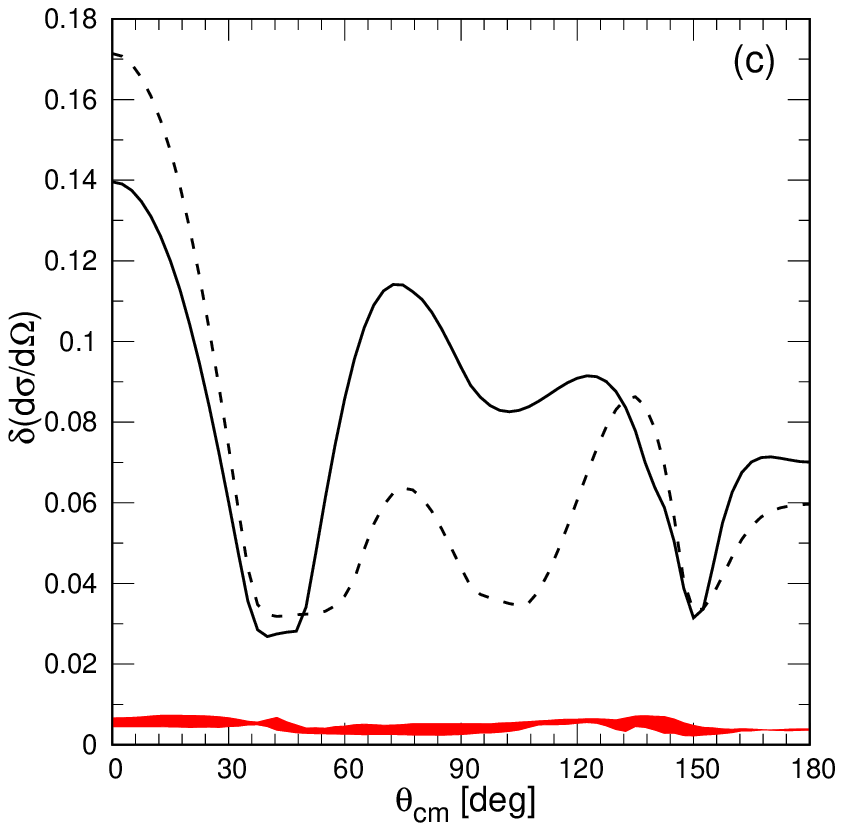}
\caption{(color online) 
The ratios $\delta(d\sigma/d\Omega) = \frac{E}{\frac{d\sigma}{d\Omega}\vert_{S_0}}$ 
with $E=\{\frac12 \Delta_{68\%}, \delta(X)^{(3)}, \delta(X)^{(3)}_{Bayes} \}$ 
that is the relative statistical uncertainties (the red band),
the relative truncation errors (Eq.~(\ref{eq_truncation})) (the solid black curve) and the relative Bayesian truncation error from the
$\bar{C}^{650}_{0.5-10}$ model (the dashed black curve),  obtained 
with the SMS N$^{2}$LO ($\Lambda =$~450~MeV) NN potential 
for the differential cross section d$\sigma$/d$\Omega$
in elastic nd scattering at the incoming nucleon laboratory energies: 
(a) $E = 13$ MeV, (b) $E = 65$ MeV and (c) $E = 135$ MeV 
as a function of the center-of-mass scattering angle $\theta_{c.m.}$. 
}
\label{fig:ds_rel_trunc_bayes_stat_N2LO}
\end{figure}

\begin{figure}
\includegraphics[width=5.3cm]{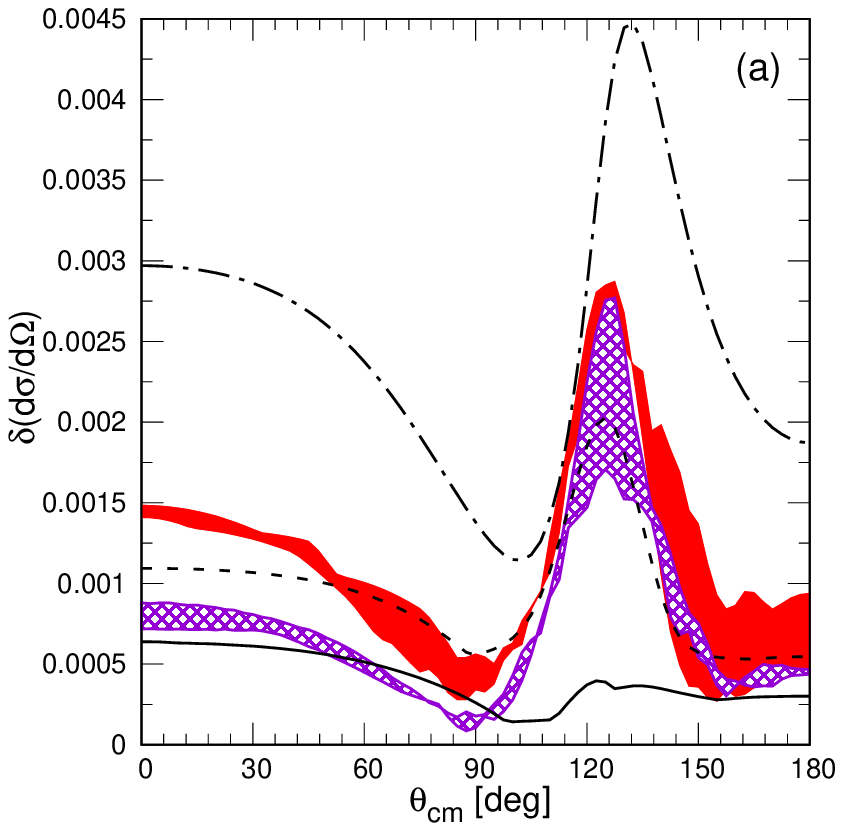}
\includegraphics[width=5.3cm]{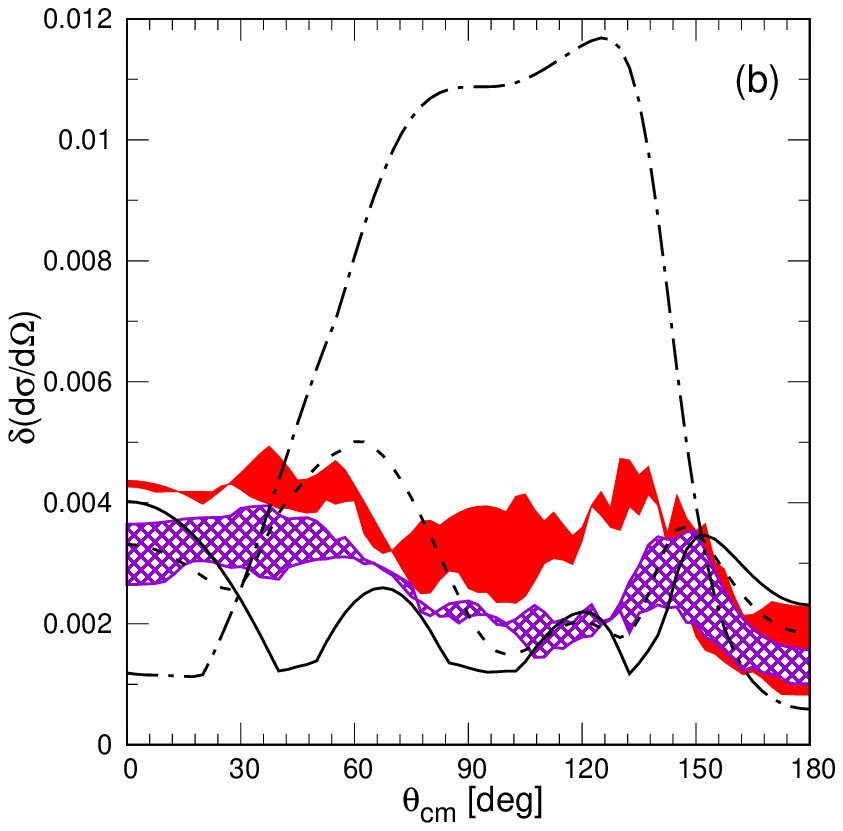}
\includegraphics[width=5.3cm]{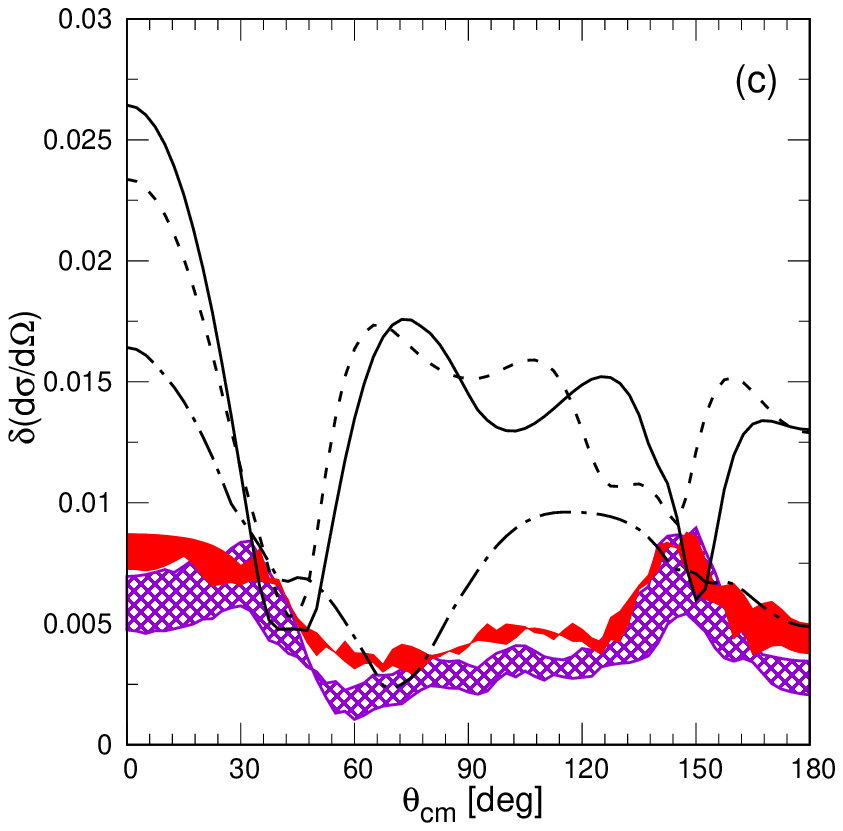}
\caption{(color online) 
The same ratios as in Fig.\ref{fig:ds_rel_trunc_bayes_stat_N2LO} but for the chiral SMS N$^{4}$LO ($\Lambda =$~450~MeV) NN potential.
The additional dark-violet band represents the relative statistical uncertainty for the OPE-Gaussian potential.
The additional black dash-dotted curve shows the ratio $\delta_{reg}$ measuring the cutoff dependence, see text.}
\label{fig:ds_rel_trunc_bayes_stat_N4LO}
\end{figure}

\begin{figure}
\includegraphics[width=5.3cm,clip=true]{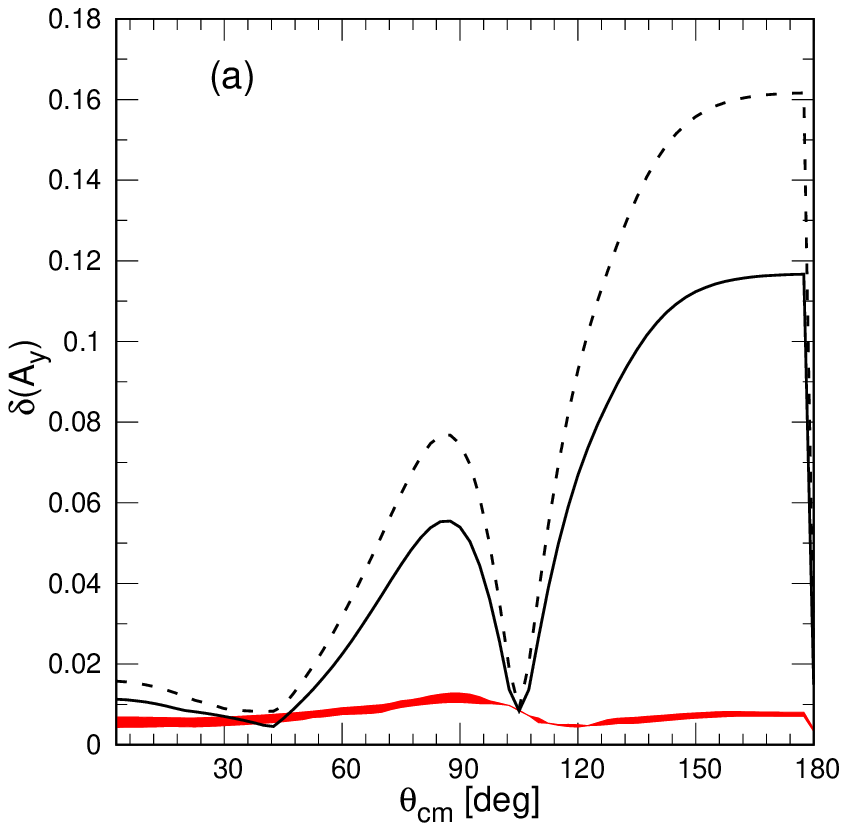}
\includegraphics[width=5.3cm,clip=true]{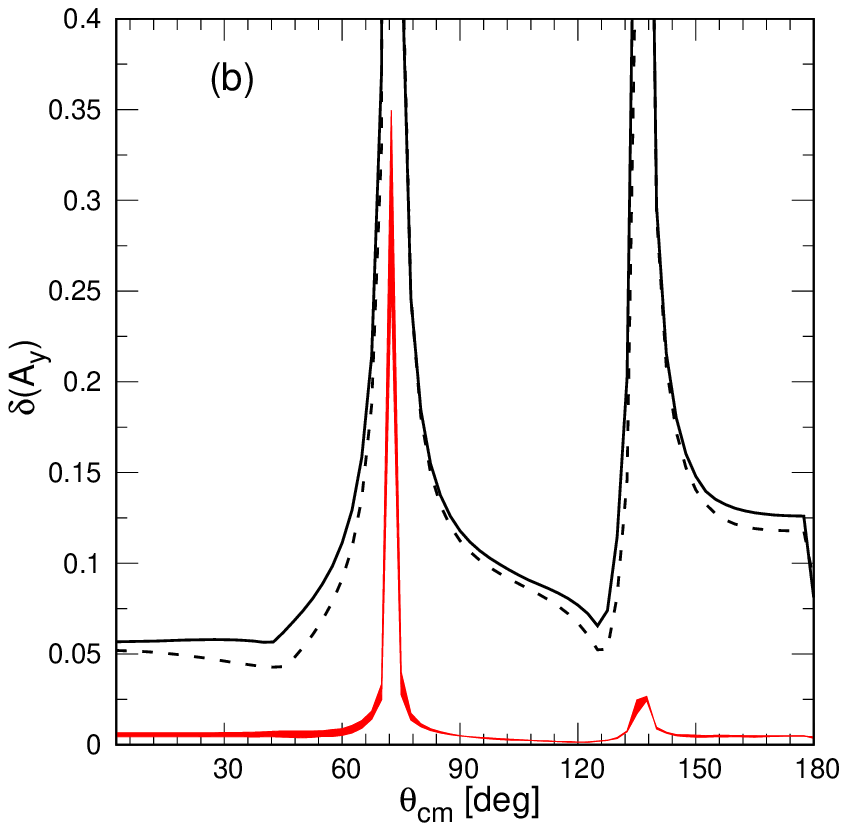}
\includegraphics[width=5.3cm,clip=true]{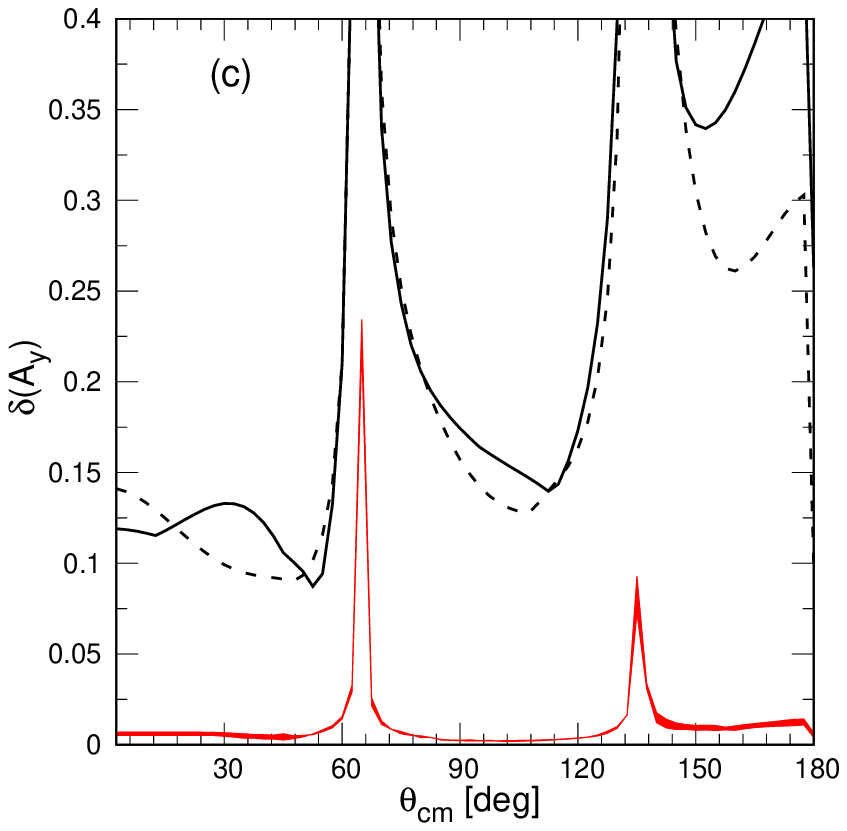}
\caption{(color online) 
The same ratios as in Fig.\ref{fig:ds_rel_trunc_bayes_stat_N2LO} but for the 
neutron analyzing power $A_{y}$. The curves and bands are the same as in Fig.~\ref{fig:ds_rel_trunc_bayes_stat_N2LO}.}
\label{fig:ay_rel_trunc_bayes_stat_N2LO}
\end{figure}

In the case of the elastic deuteron vector analyzing power iT$_{11}$ (Fig.~\ref{fig5}) the relative difference 
between the widths of two bands of predictions i.e. $\vert \frac12 \Delta_{68\%} - \delta(X)^{(5)}\vert / (\frac{1}{2}(\frac12 \Delta_{68\%} + \delta(X)^{(5)}))$ 
at $E =$~65~MeV reaches a few percent at scattering angle $\theta_{c.m.}$=~90$^{\circ}$. However, with increasing energy there is a significant increase in the magnitude of the truncation error which leads to an increase of the relative difference between the widths of two bands. 
For instance, at $E =$~135~MeV and $\theta_{c.m.} =$~90$^{\circ}$, that difference approaches
about 84$\%$ (with $\Delta^{(5)} > \frac12 \Delta_{68\%}$), but already at $E =$~200~MeV it amounts to 92$\%$. 
Similarly, for the deuteron to neutron spin-transfer coefficient $K^{y}_{y'}$(d-n), we observe 
at $\theta_{c.m.}$=~90$^{\circ}$ that at $E =$~65~MeV the difference between the statistical and truncation errors is 
almost invisible, but at $E = 135$~MeV and $E =$~200~MeV it amounts up to 20$\%$ and 82$\%$, respectively. 
Last but not least, we have to note that the ratios of the
magnitude of the statistical uncertainties to the magnitude of the truncation error, that is 
$\frac{\frac12 \Delta_{68\%}}{\delta(X)^{(5)}}$,
for the polarization observables are for most of the scattering angles much bigger than the same ratios but for 
the cross section. Probably this is due to a bigger sensitivity of polarization observables to the specific 
partial wave potential parameters of the chiral interaction used.

In Tab.~\ref{tab1} we give details on the statistical uncertainties and the truncation  errors for the deuteron vector analyzing power iT$_{11}$
shown in Fig.~\ref{fig5}. Here, beside the predictions for iT$_{11}$ obtained with the SMS N$^4$LO potential we 
also show the magnitudes of the statistical uncertainties ($\frac12 \Delta_{68\%}$) and truncation errors ($\delta(X)^{(5)}$).
Again, the rapid decrease of $\frac{\Delta_{68\%}}{\delta(X)^{(5)}}$ with the energy can be observed.
The predictions based on the genuine set of the potential parameters $S_0$, shown in the third column of Tab.~\ref{tab1} 
do not need to be in the centre of predictions obtained with various sets of the potential parameters.
Thus in the 4th and the 5th columns of Tab.~\ref{tab1} we give distances between the predictions from the 
3rd column and minimal and maximal predictions among those based on 34 sets of potential parameters taken
into account when calculating $\Delta_{68\%}$. The different magnitudes of these distances, at the given energy and scattering angle,
point to a nonlinear dependence of the 3N observables on the NN potential parameters. 

\begin{figure}
\includegraphics[width=5.3cm]{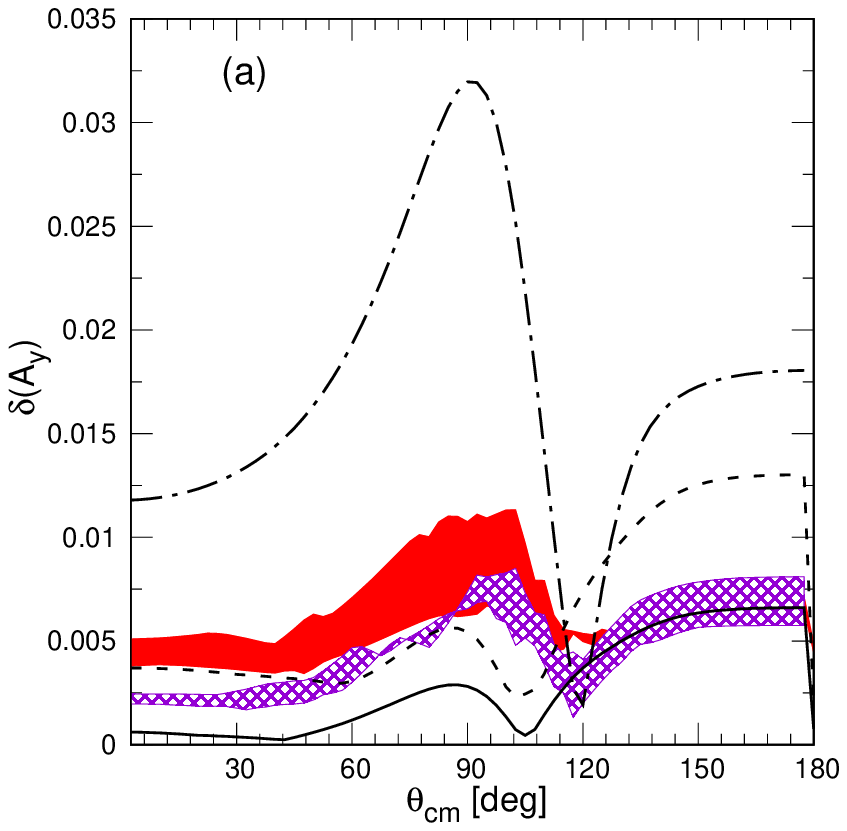}
\includegraphics[width=5.3cm]{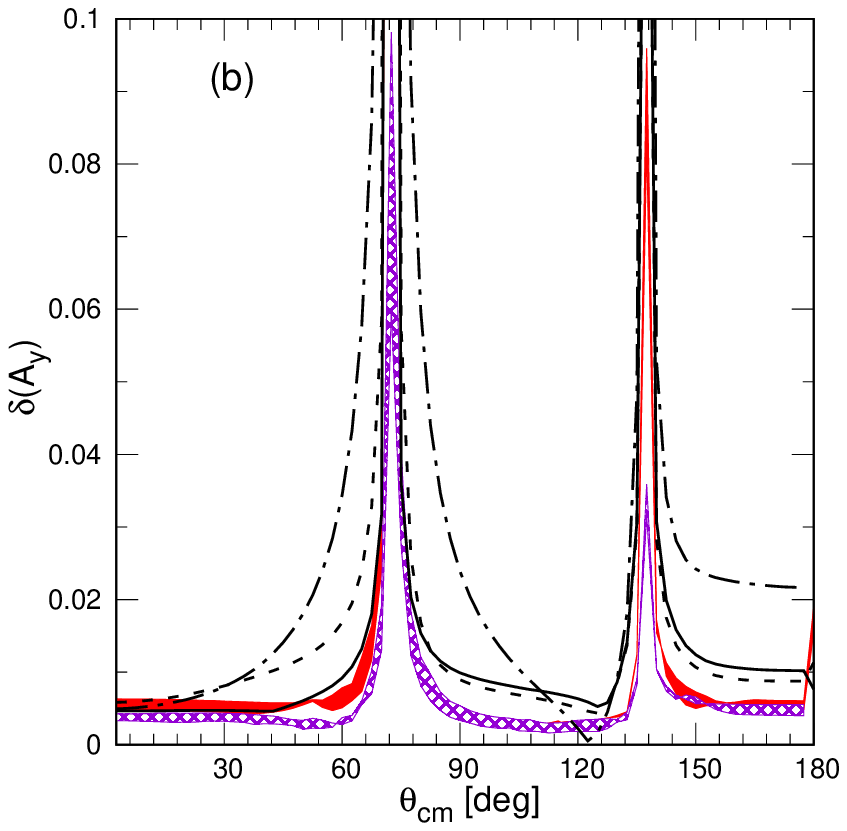}
\includegraphics[width=5.3cm]{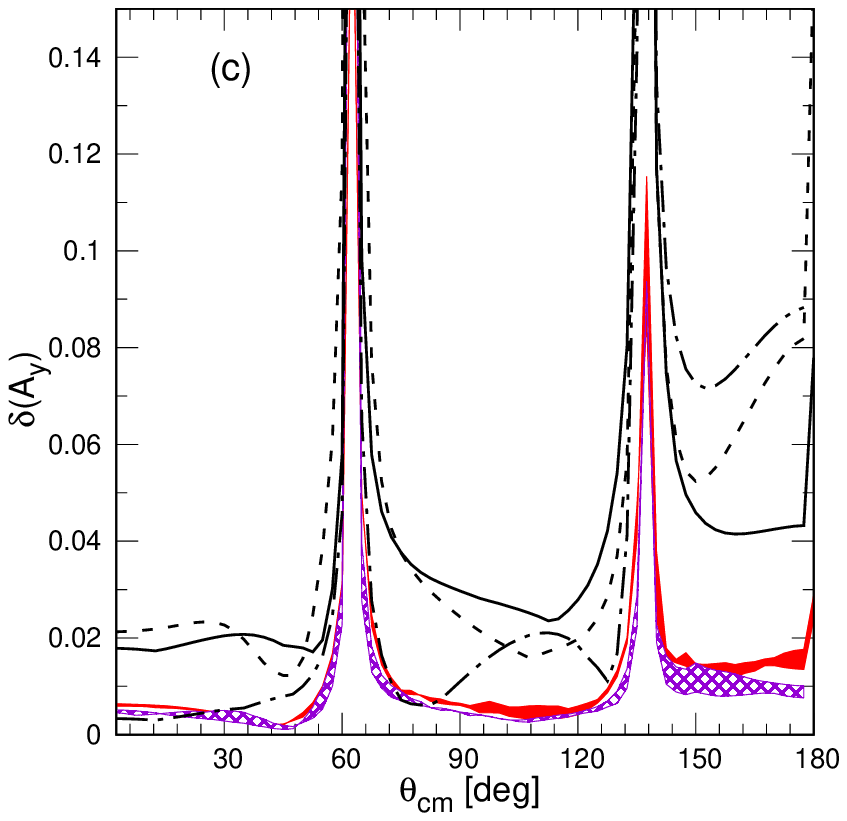}
\caption{(color online) 
The same ratios as in Fig.\ref{fig:ds_rel_trunc_bayes_stat_N4LO} but for the 
neutron analyzing power $A_{y}$. The curves and bands are as in Fig.~\ref{fig:ds_rel_trunc_bayes_stat_N4LO}
}
\label{fig:ay_rel_trunc_bayes_stat_N4LO}
\end{figure}

Bayesian statistics also yields a general and statistically well-founded approach to quantify 
truncation errors in perturbative calculations. 
We employ here the same Bayesian procedure as already used by the LENPIC Collaboration 
to study truncation errors in NN and 3N scattering~\cite{Epelbaum1-arxiv}, which is a slightly modified 
version of the Bayesian approach developed in Refs.~\cite{Furnstahl, Melendez}. Therefore, in the following we again only 
briefly describe our Bayesian procedure to determine the truncation errors and 
focus on a comparison of its results with the previously discussed statistical and truncation errors.

Rewriting Eq.~(\ref{eq_truncation0}) in terms of dimensionless expansion coefficients $c_{i}$ in the form
\begin{equation}
X = X_{ref}\left( c_{0} + c_{2}Q^{2} + c_{3}Q^{3} + c_{4}Q^{4} + \ldots\right)\;, 
\label{eq_truncation1}
\end{equation}
setting the overall scale $X_{ref}$, with $\Delta X^{(i)}$ given in Eq.~(\ref{eq_truncation}), as
\begin{equation}
  X_{ref} =
  \begin{cases}
    max\left(\vert X^{(0)}\vert, Q^{-2} \vert \Delta X^{(2)}\vert \right)  & \text{for $k = 2$}\;, \\
    max\left(\vert X^{(0)}\vert, Q^{-2} \vert \Delta X^{(2)}\vert, Q^{-3} \vert \Delta X^{(3)}\vert \right) & \text{for $k \geq 3$}\;,
  \end{cases}
\end{equation}
and assuming that $\Delta X^{(i)}$ are known explicitly up to the order $X^{(k)}, k\geq 2$, one 
can estimate the size of the truncation error at the k-th order of the chiral expansion as $\delta X^{(k)}_{Bayes} \equiv X_{ref}\Delta$ where
$\Delta \equiv \sum^{\infty}_{i = k + 1}c_{i}Q^{i} \approx \sum^{k + h}_{i = k + 1}c_{i}Q^{i}$ is distributed, 
given the knowledge of $\left\lbrace c_{i \leq k}\right\rbrace$
with a posterior probability density function
\begin{equation}
\text{pr}_{h}\left(\Delta\mid \left\lbrace c_{i \leq k} \right\rbrace \right) = \frac{\int^{\infty}_{0}d\bar{c}~\text{pr}_{h}\left(\Delta\mid\bar{c} \right)\text{pr}(\bar{c})\prod_{i \in A}\text{pr}(c_{i}\mid\bar{c})}{\int^{\infty}_{0}d\bar{c}~\text{pr}(\bar{c})\prod_{i \in A}\text{pr}(c_{i}\mid\bar{c})}\;.
\label{eq_posterior}
\end{equation}
Here the prior probability density function $\text{pr}(c_{i}\mid\bar{c})$ is taken in the form of the Gaussian N(0,$\bar{c}^2$) function 
and $\text{pr}(\bar{c})$ is a log-uniform distribution in the range $(\bar{c}_{<},\bar{c}_{>})$.
Set $A$ is defined as 
$A = \left\lbrace n \in \mathbb{N}_{0} \mid n \leq k~\wedge~n \neq 1 \wedge n \neq m \right\rbrace $, $m \in  \left\lbrace 0, 2, 3 \right\rbrace$ and 
\begin{equation}
\text{pr}_{h}(\Delta\mid\bar{c}) \equiv \left[\prod^{k + h}_{i = k + 1}\int^{\infty}_{-\infty}dc_{i}\text{pr}(c_{i}\mid\bar{c}) \right] \delta\left[\Delta - \sum^{k + h}_{j = k + 1}c_{j}Q^{j} \right]\;,
\end{equation}
with $h$ being the number of the chiral orders above $k$ which contribute to the truncation error.
The resulting $\text{pr}_{h}(\Delta\mid \left\lbrace c_{i \leq k} \right\rbrace)$ is symmetric with respect to $\Delta = 0$ so
one can find the degree-of-belief (DoB) interval $(-d^{(p)}_{k},d^{(p)}_{k})$  at the probability $p$, 
as a solution to the inverse problem given by the numerical integration
\begin{equation}
p = \int_{-d^{(p)}_{k}}^{d^{(p)}_{k}} \text{pr}_{h}(\Delta\mid \left\lbrace c_{i \leq k} \right\rbrace) d\Delta
\end{equation}
and consequently the truncation error $\delta X^{(k)}_{Bayes}=X_{ref} d^{(p)}_{k}$.
In the following we use $h=10$, $\bar{c}_{<}=0.5$, $\bar{c}_{>}=10$, $\Lambda_{b}=650$~MeV and $M^{eff}_{\pi}=200$~MeV.
The two latter quantities enter the expansion parameter
$Q = max\left( \frac{p}{\Lambda_{b}}, \frac{M^{eff}_{\pi}}{\Lambda_{b}}\right)$ with momentum scale $p$ defined in Eq.(17) of Ref.~\cite{Epelbaum1-arxiv}.
The detailed expression for $\text{pr}_{h}\left(\Delta\mid \left\lbrace c_{i \leq k} \right\rbrace \right)$ for assumed priors can be found
in Appendix A of Ref.~\cite{Epelbaum1-arxiv} and our choice of the $h, \bar{c}_{<}, \bar{c}_{>}$ and $\Lambda_{b}$ values corresponds to the model
$\bar{C}^{650}_{0.5-10}$ from Ref.~\cite{Epelbaum1-arxiv}.

In Fig.~\ref{bayes_stat_N2LO} we show the differential cross section and the neutron analyzing 
power $A_{y}$ in elastic neutron-deuteron scattering at N$^{2}$LO at the laboratory energies $E=$13, 65 and 135~MeV 
for the cutoff value $\Lambda =$~450~MeV, along with the truncation error corresponding to the 68\% DoB interval 
and the statistical uncertainty obtained with the same force. For the differential cross 
section both types of errors almost overlap at $E =$ 13 MeV but with the increasing energy the magnitude 
of 68\% DoB interval from the $\bar{C}^{650}_{0.5-10}$ Bayesian model exceeds the statistical uncertainty at forward 
and backward scattering angles as well as at the minimum of the cross section. This is more noticeable for $A_{y}$. In this case the truncation error 
proves to be
much bigger than the statistical uncertainty at all energies. This domination of truncation errors appears in specific ranges of the scattering angle
for two lower energies and at the $E$=135~MeV the truncation errors exceed the statistical ones in the whole angular domain.
  
To facilitate more insight into the magnitudes of the theoretical uncertainties
we compute the ratios of the theoretical errors and the predictions based on the genuine set of the potential parameters (set $S_0$).
They are presented in Figs.~\ref{fig:ds_rel_trunc_bayes_stat_N2LO}-\ref{fig:ay_rel_trunc_bayes_stat_N4LO}. 
Fig.~\ref{fig:ds_rel_trunc_bayes_stat_N2LO} confirms findings from Fig.~\ref{bayes_stat_N2LO} (and from Figs.~\ref{fig5}-\ref{fig6}
for other observables and at N$^4$LO), that the magnitude of the statistical uncertainty is much smaller than the 
truncation errors obtained within both methods. In the case of the chiral SMS N$^{4}$LO potential shown in Fig.~\ref{fig:ds_rel_trunc_bayes_stat_N4LO} 
one observes more complex relations between the two types of the relative errors. 
As can be seen in Figs.~\ref{fig:ds_rel_trunc_bayes_stat_N4LO}(a) and ~\ref{fig:ds_rel_trunc_bayes_stat_N4LO}(b), 
for the differential cross section at $E =$~13 MeV and at $E =$~65 MeV, the magnitude of the statistical uncertainty 
is bigger or comparable to the magnitude of the truncation errors computed in the two approaches. 
In the same figure we show also the uncertainty 
due to using various values of the cutoff parameter when regularizing the potential, which is related to the truncation uncertainty discussed above.
We define it for the observable $X$ as $\delta_{reg} \equiv \frac{ \frac12 ({\rm max}_j(X_j)-{\rm min}_j(X_j))}{X_2}$, where 
the subscript $j \in \{ 1,2,3,4 \}$
corresponds to different values of the cutoff parameter $\Lambda$=400, 450, 500, and 550 MeV, respectively.
For both lower energies the uncertainty related to the cutoff parameters is much bigger than the remaining theoretical errors.
At $E$= 135 MeV the relative errors for statistical uncertainties are smaller compared to the truncation ones.
Thus we observe that at this energy the truncation errors become a dominant source of the total theoretical uncertainty
for calculations within a given chiral force. This situation will likely change after applying higher-order
contributions to the NN chiral force, what should reduce the truncation error. The magnitude of the truncation errors is, as expected,
much smaller at N$^4$LO than at N$^2$LO and the magnitude of the statistical uncertainties remains similar 
at these two orders of the chiral expansion and at the same reaction energy. 
In Fig.~\ref{fig:ds_rel_trunc_bayes_stat_N4LO}
the statistical uncertainty for the chiral SMS N$^{4}$LO results is also compared with the results based on the OPE-Gaussian force.
The latter is slightly smaller than the statistical uncertainty of the chiral prediction in the whole range 
of the scattering angle. It is worth noting that the absolute values of the relative errors remain below
0.5\%, 1.2\% and 3\% at $E$=13, 65 and 135 MeV, respectively. This proves the high quality of the SMS potential at N$^4$LO 
and the reliability of predictions based on this interaction.

The relative statistical uncertainty from the chiral SMS N$^{2}$LO potential is smaller than the relative truncation 
errors for the neutron analyzing power $A_{y}$ at all three energies presented 
in Fig.~\ref{fig:ay_rel_trunc_bayes_stat_N2LO}. This picture is similar to the one for the differential cross section 
with the same interaction, shown in Fig.~\ref{fig:ds_rel_trunc_bayes_stat_N2LO}. 
Increasing order of the chiral expansion to N$^{4}$LO, see Fig.~\ref{fig:ay_rel_trunc_bayes_stat_N4LO}, 
the magnitude of the relative statistical uncertainty for $A_{y}$ hardly changes.
It is also similar to the magnitude of the same ratio for the OPE-Gaussian-potential-based predictions.
At $E =$~13 MeV and below  $\theta_{c.m.} \approx 115^{\circ}$ the relative statistical error again 
is bigger than the relative truncation uncertainty. The truncation errors grow significantly for both higher 
energies 
as displayed in Figs.~\ref{fig:ay_rel_trunc_bayes_stat_N4LO}(b) and ~\ref{fig:ay_rel_trunc_bayes_stat_N4LO}(c).
The uncertainty related to the cutoff dependence, important at low and medium energies, is surpassed by the truncation errors
at $E$=135 MeV. The magnitudes of all the types of the uncertainties for $A_{y}$ at $E$=13, 65, and 135 MeV 
remain below approx. 1.5\%, 2\%, and 4\%, except for regions of the scattering angle where $A_{y}$ reaches zero.

The estimation of theoretical uncertainty shown in this section bases on predictions 
of only NN interaction which are incomplete from the third order of the chiral expansion, where the three-nucleon interaction 
starts to contribute. It is very likely that the estimated truncation errors will change after inclusion of the 3NF.
This should be tested as soon as a 3NF consistent with the SMS NN potential is available. 

\section{\label{sec:ResultBreakup}Results for the deuteron breakup reaction}
\label{Sec_Break}

In the case of the neutron induced deuteron breakup reaction, we have selected a few kinematical complete
configurations to exemplify only the statistical uncertainties for observables in this process.
\begin{figure}[ht]
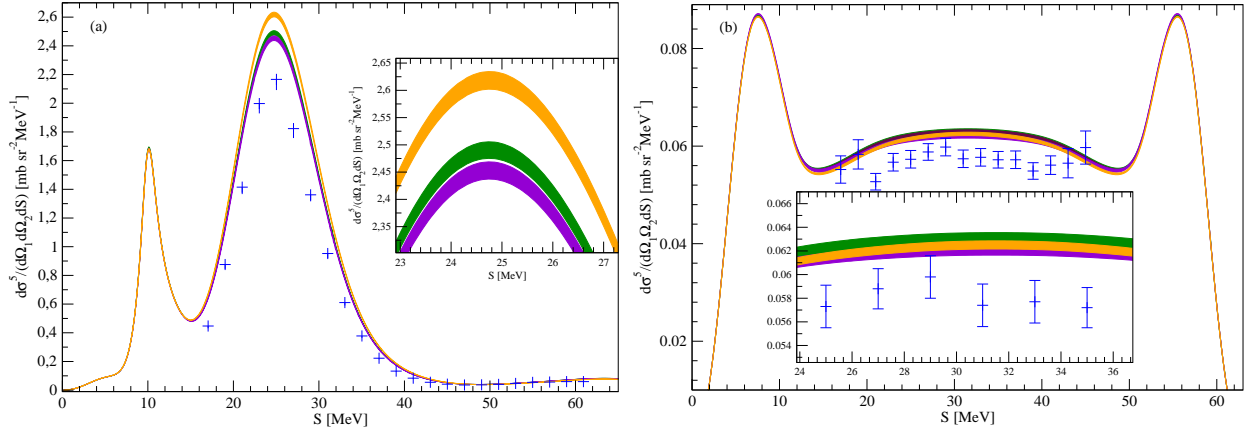

\begin{minipage}[h]{0.495\linewidth}
\center{\includegraphics[width=1.\textwidth,clip=true]{break_e65_d5sigma_s_30p0_0p0_59p5_180p0.eps}} \\
\end{minipage}
\hfill
\begin{minipage}[h]{0.495\linewidth}
\center{\includegraphics[width=1.\textwidth,clip=true]{break_e65_d5sigma_s_54p0_0p0_54p0_120p0.eps}}  
\end{minipage}
\caption{The five-fold cross section $\frac{d^{5}\sigma}{d\Omega_{1}d\Omega_{2}dS}$ for the $d(n,n_{1}n_{2})p$ 
breakup reaction at the incoming nucleon laboratory energy $E$=~65~MeV shown as a function of the arc-length $S$ for the following 
polar angles $\theta_{i}$ and the relative azimuthal angle $\phi_{12}$ of the momenta of two detected neutrons:
(a) $\theta_{1} = 30.5^{\circ}, \theta_{2} = 59.5^{\circ}, \phi_{12} = 180^{\circ}$ (QFS configuration) 
and (b) $\theta_{1} = \theta_{2} = 54.0^{\circ}, \phi_{12} = 120^{\circ}$ (SST configuration). 
The orange, green and violet bands represent statistical uncertainties obtained with the OPE-Gaussian force, the chiral N$^{4}$LO and N$^{4}$LO$^+$ ($\Lambda~=~450$~MeV), respectively. The experimental proton-deuteron data are from Ref.~\cite{Allet} for (a) and from Ref.~\cite{Zejma} for (b).}
\label{fig10} 
\end{figure}

\begin{figure}[ht]
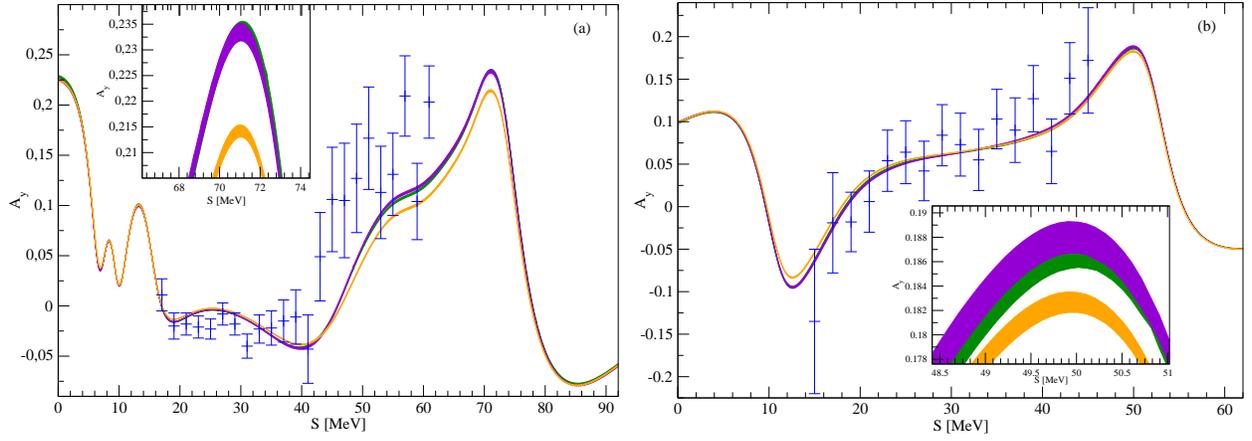

\begin{minipage}[h]{0.495\linewidth}
\center{\includegraphics[width=1.\textwidth,clip=true]{break_e65_ayn_s_30p0_0p0_59p5_180p0.eps}} \\
\end{minipage}
\hfill
\begin{minipage}[h]{0.495\linewidth}
\center{\includegraphics[width=1.\textwidth,clip=true]{break_e65_ayn_s_54p0_0p0_54p0_120p0.eps}}  
\end{minipage}
\caption{The neutron analyzing power $A_{y}$(n) for the $d(n,n_{1}n_{2})p$ breakup reaction at the incoming nucleon laboratory 
energy $E$=~65~MeV 
for the following polar angles and the relative azimuthal angle of the momenta of two detected neutrons:
(a) $\theta_{1} = 30.5^{\circ}, \theta_{2} = 59.5^{\circ}, \phi_{12} = 180^{\circ}$ and (b) $\theta_{1} = \theta_{2} = 54.0^{\circ}, \phi_{12} = 120^{\circ}$. The curves and bands as in Fig.~\ref{fig10}. The proton-deuteron data are from Ref.~\cite{Allet}.}
\label{fig11} 
\end{figure}

Proceeding in the same way as for elastic nd scattering, 
we estimate the theoretical statistical uncertainties of nd breakup observables, due to uncertainty of the SMS NN potential parameters. 
We show in Fig.~\ref{fig10} these uncertainties for the neutron-induced deuteron breakup cross section obtained with the chiral SMS potential 
with $\Lambda =$~450~MeV, at two orders of the chiral expansion, N$^4$LO and N$^4$LO$^+$,  and compare 
them with the corresponding results obtained with the OPE-Gaussian interaction. 
The magnitudes of the statistical uncertainties for the cross section 
reach their maximum approximately at $S =$~25~MeV for the quasi-free scattering (QFS) configuration in Fig.~\ref{fig10}a. 
Predictions obtained with the chiral N$^{4}$LO and N$^{4}$LO$^+$ potentials (at $\Lambda=$~450~MeV) differ slightly each other
but the OPE-Gaussian force based results are clearly separated from the two chiral predictions. 
For the space-star configuration (SST) (Fig.~\ref{fig10}b) the predictions of three potentials practically overlap.
\begin{figure}[h]
\includegraphics[width=1\textwidth,clip=true]{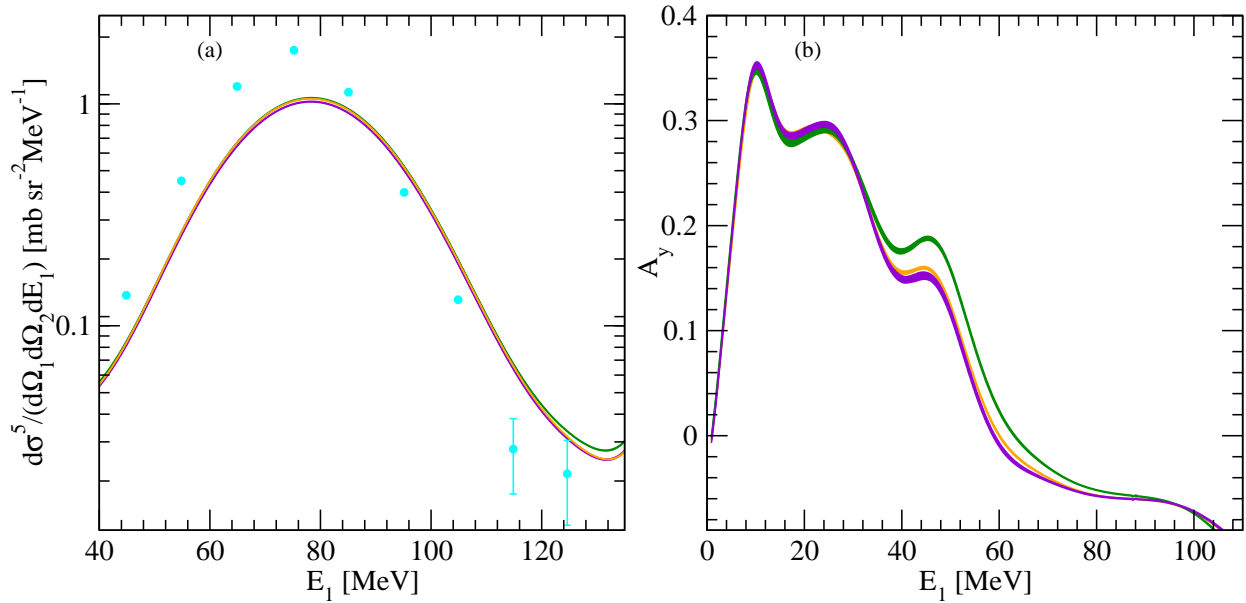}
\caption{(color online) (a) The five-fold differential cross section $\frac{d^{5}\sigma}{d\Omega_{1}d\Omega_{2}dE_{1}}$ 
as a function of the laboratory kinetic energy $E_1$ of the first detected nucleon  
for the following detection angles of two neutrons:
$\theta_{1} = 45.0^{\circ}, \theta_{2} = 35.0^{\circ}, \phi_{12} = 180^{\circ}$ and (b) the neutron analyzing power for the $d(n,n_{1}n_{2})p$ breakup reaction 
as a function of the laboratory kinetic energy $E_1$ of the first detected nucleon
for the following detection angles of two neutrons $\theta_{1} = 52.0^{\circ}, \theta_{2} = 45.0^{\circ}, \phi_{12} = 180^{\circ}$. The incoming neutron laboratory kinetic energy is $E$=200~MeV. 
The curves and bands are as in Fig.~\ref{fig10}. The experimental proton-deuteron data in (a) are from Ref.~\cite{Pairsuwan}.}
\label{fig12} 
\end{figure}
In Figs.~\ref{fig11}a and~\ref{fig11}b we exemplify the neutron analyzing power for the QFS and SST 
configurations at $E =$~65~MeV, respectively. Here the statistical uncertainties remain negligible for both configurations. The differences between predictions based on the OPE-Gaussian force and the chiral potentials at N$^4$LO and N$^4$LO$^+$ amount up to 7$\%$ as seen in the maximum of the A$_{y}$(n) for the SST configuration.
Fig.~\ref{fig12} exemplifies that at the higher energy $E =$~200~MeV the statistical uncertainties remain small. 
It is also interesting to note that for the breakup process there exist kinematical configurations for which 
a clear difference between chiral predictions at N$^4$LO and N$^4$LO$^+$ is observed. 
This is exemplified in Fig.~\ref{fig12}b, where the difference between 
results for the nucleon analyzing power around $E_1$=48~MeV 
at these two orders of the chiral expansion reaches $\approx 20\%$.
  
\section{\label{sec:Summary}Summary and conclusions}
\label{Summary}
We employed the new high-quality $\chi$EFT NN potential with the semi-local regularization in momentum space at different orders of chiral expansion up to N$^{4}$LO$^+$ to describe the elastic nd scattering and the neutron-induced breakup reactions at energies up to 200~MeV.
We used the correlation matrix of that NN potential parameters to study the propagation of uncertainties from the NN potential parameters to 3N observables. Next we compared these uncertainties with the truncation errors estimated using two different approaches: the prescription from Ref.~\cite{lenpic4} 
and the Bayesian approach from Refs.~\cite{Epelbaum1-arxiv, Furnstahl, Melendez}. 
We calculated also the uncertainty of predictions induced by different values of the regularization cutoff parameter used.

The description of the data delivered by the chiral force with the semi-local momentum-space regularization is similar to that
based on the older versions of the chiral potential from the Bochum-Bonn group. 
Our findings confirm that the statistical uncertainties of the elastic nd scattering observables are smaller than the dispersion of results 
arising from using various orders of chiral NN interactions, both at low- and at high-energies. 
We find that statistical errors remain still relatively small in the deuteron breakup process
at the considered kinematical configuration independently from the employed NN force model. 
The statistical uncertainties of the chiral predictions have similar magnitudes and the energy dependence
as those from the semi-phenomenological OPE-Gaussian force.

Clearly, at low and medium energies the regulator dependence dominates other types of uncertainties.
Also the truncation errors found in our studies are not negligible.
Only at low energies and at N$^4$LO truncation errors become smaller than statistical uncertainties, 
both for the cross section and the neutron analyzing power. 
However, the estimated magnitudes of all types of uncertainties remain small, usually in the range 0.5\%-4\%,
depending on the energy and the observable.
The fact that various contributions to the theoretical 
uncertainty are so small points to the high quality of the theoretical input in the SMS interaction.

Summarizing, our analysis of theoretical uncertainties in the neutron-deuteron scattering 
confirms the SMS chiral potential belongs to the first-rate models of nuclear forces. It also demonstrates that, with an ongoing  
progress in the derivation, regularization and inclusion of higher-order contributions to the nuclear interaction, 
theoretical uncertainties, obtained with the chiral interaction, would be reduced to the limit 
dependent only on the quality of experimental data which influence the statistical errors. 
Presently this is observed for a fixed value of the regulator parameter only at low energies 
but very likely this region will be extended to much higher energy values.

\acknowledgments
This work is a part of the LENPIC project and was supported by the Polish National Science Centre under Grants No. 2016/22/M/ST2/00173 and No. 2016/21/D/ST2/01120. 
It was also supported in part by BMBF (Grant No. 05P18PCFP1) and by DFG through funds provided to the Sino-German CRC 110 “Symmetries and the Emergence
of Structure in QCD” (Grant No. TRR110). Numerical calculations were performed on the supercomputer cluster of the JSC, J\"ulich, Germany.

\end{document}